\documentclass[aps,prd,twocolumn,amsfonts,amssymb,amsmath,showpacs,letterpaper,superscriptaddress,nofootinbib,altaffilletter,fontenc]{revtex4-1}

\usepackage[citecolor=blue,linkcolor=blue,colorlinks=true,urlcolor=blue]{hyperref}

\usepackage{color}
\usepackage{graphicx}
\usepackage{latexsym}
\usepackage{float}
\usepackage{amsmath}
\usepackage{amssymb}
\usepackage{multirow}
\usepackage{ifthen}
\usepackage{slashed}
\usepackage{lineno}
\usepackage{makecell} 
\usepackage[mathscr]{euscript} 
\usepackage{mathrsfs} 
\usepackage{subfigure}

\usepackage{longtable}
\usepackage{dcolumn}
\usepackage{array}
\usepackage{ragged2e}
\usepackage{multirow}
\usepackage[utf8]{inputenc}

\usepackage{tabularx,booktabs}
\newcolumntype{Y}{>{\centering\arraybackslash}X}
\usepackage{booktabs}
\usepackage{pifont}

\usepackage{array}

\newcolumntype{R}[1]{>{\RaggedRight}p{#1}}

\usepackage[normalem]{ulem}

\renewcommand{\today}{\number\day\space\ifcase\month\or
  January\or February\or March\or April\or May\or June\or
  July\or August\or September\or October\or November\or December\fi
  \space\number\year}

\def\be{\begin{equation}}
\def\ee{\end{equation}}
\def\bi{\begin{itemize}} 
\def\ei{\end{itemize}}
\def\ben{\begin{enumerate}}
\def\een{\end{enumerate}}

\def\apj{Astrophys. J.}

\def\aap{Astron. Astrophys. }

\def\prl{Phys. Rev. Lett.}
\def\prd{Phys. Rev. D.}

\begin{document}


\title{Detecting and reconstructing gravitational waves from the next Galactic core-collapse supernova in the Advanced Detector Era
}

\author{Marek~J.~Szczepa\'nczyk}
\email[E-mail: ]{marek.szczepanczyk@ligo.org}
\affiliation{University of Florida, Gainesville, FL 32611, USA}
\author{Javier~M.~Antelis}
\affiliation{Embry-Riddle Aeronautical University, Prescott, AZ 86301, USA}
\affiliation{Tecnologico de Monterrey, Escuela de Ingeniería y Ciencias, Monterrey, N.L., 64849, México}
\author{Michael~Benjamin}
\affiliation{Embry-Riddle Aeronautical University, Prescott, AZ 86301, USA}
\author{Marco~Cavagli\`a}
\affiliation{Institute of Multi-messenger Astrophysics and Cosmology, Missouri University of Science and Technology, Rolla, MO 65409, USA}
\author{Dorota~Gondek-Rosi\'nska}
\affiliation{Astronomical Observatory Warsaw University, 00-478 Warsaw, Poland}
\author{Travis~Hansen}
\affiliation{Embry-Riddle Aeronautical University, Prescott, AZ 86301, USA}
\author{Sergey~Klimenko}
\affiliation{University of Florida, Gainesville, FL 32611, USA}
\author{Manuel~D.~Morales}
\affiliation{Universidad de Guadalajara, Guadalajara, Jal., 44430, M\'exico}
\author{Claudia~Moreno}
\affiliation{Embry-Riddle Aeronautical University, Prescott, AZ 86301, USA}
\affiliation{Universidad de Guadalajara, Guadalajara, Jal., 44430, M\'exico}
\author{Soma~Mukherjee}
\affiliation{The University of Texas Rio Grande Valley, Brownsville, TX 78520, USA}
\author{Gaukhar~Nurbek}
\affiliation{The University of Texas Rio Grande Valley, Brownsville, TX 78520, USA}
\author{Jade~Powell}
\affiliation{Centre for Astrophysics and Supercomputing, Swinburne University of Technology, Hawthorn, VIC 3122, Australia}
\affiliation{ARC Centre of Excellence for Gravitational Wave Discovery (OzGrav), Melbourne, Australia}
\author{Neha~Singh}
\affiliation{Astronomical Observatory Warsaw University, 00-478 Warsaw, Poland}
\author{Satzhan~Sitmukhambetov}
\affiliation{The University of Texas Rio Grande Valley, Brownsville, TX 78520, USA}
\author{Pawe\l~Szewczyk}
\affiliation{Astronomical Observatory Warsaw University, 00-478 Warsaw, Poland}
\author{Jonathan~Westhouse}
\affiliation{Embry-Riddle Aeronautical University, Prescott, AZ 86301, USA}
\author{Oscar~Valdez}
\affiliation{The University of Texas Rio Grande Valley, Brownsville, TX 78520, USA}
\author{Gabriele~Vedovato}
\address {Universit\`a di Padova, Dipartimento di Fisica e Astronomia, I-35131 Padova, Italy }
\address {INFN, Sezione di Padova, I-35131 Padova, Italy }
\author{Yanyan~Zheng}
\affiliation{Institute of Multi-messenger Astrophysics and Cosmology, Missouri University of Science and Technology, Rolla, MO 65409, USA}
\author{Michele~Zanolin}
\affiliation{Embry-Riddle Aeronautical University, Prescott, AZ 86301, USA}


\begin{abstract}

We performed a detailed analysis of the detectability of a wide range of gravitational waves derived from core-collapse supernova simulations using gravitational-wave detector noise scaled to the sensitivity of the upcoming fourth and fifth observing runs of the Advanced LIGO, Advanced Virgo, and KAGRA. We use the coherent WaveBurst algorithm, which was used in the previous observing runs to search for gravitational waves from core-collapse supernovae. As coherent WaveBurst makes minimal assumptions on the morphology of a gravitational-wave signal, it can play an important role in the first detection of gravitational waves from an event in the Milky Way. We predict that signals from neutrino-driven explosions could be detected up to an average distance of 10\,kpc, and distances of over 100\,kpc can be reached for explosions of rapidly rotating progenitor stars. An estimated minimum signal-to-noise ratio of 10--25 is needed for the signals to be detected. We quantify the accuracy of the waveforms reconstructed with coherent WaveBurst and we determine that the most challenging signals to reconstruct are those produced in long-duration neutrino-driven explosions and models that form black holes a few seconds after the core bounce.

\end{abstract}

\date[\relax]{Dated: \today }


\maketitle


\section{Introduction}

The discovery of gravitational waves (GWs) from a binary black hole merger~\cite{Abbott:2016blz} marked the beginning of GW astronomy. Similarly, the discovery of a binary neutron star in both GW and electromagnetic spectra~\cite{TheLIGOScientific:2017qsa} began the era of multimessenger astronomy with GWs. While the first, second, and third observing runs (O1, O2, O3) brought a wealth of binary coalescence discoveries~\cite{LIGOScientific:2018mvr,Abbott:2020niy}, we expect these numbers to grow with upcoming detectors upgrades. Currently, all detected sources are binary systems and we are waiting for a short-duration GW transient (burst). The most prominent source is a core-collapse supernova (CCSN).

CCSNe are the violent explosions of massive stars (above 8\,$M_\odot$) and are believed to form most of the black holes (BHs) detected by Advanced LIGO~\cite{TheLIGOScientific:2014jea} and Advanced Virgo~\cite{TheVirgo:2014hva}. Despite the growing understanding of stellar collapse, the explosion mechanism is not yet fully understood~\cite{Janka:2012wk}. All supernovae known to date were detected electromagnetically and low energy neutrinos were observed from only  SN~1987A~\cite{hirata:87,bionta:87,1987ESOC...26..237A}.  Unfortunately, the measured light is emitted after the initial collapse, losing all the detailed information. To understand CCSNe we need to be able to directly probe their inner dynamics. The neutrinos and GWs leave the core around the collapse time and they can be used to directly probe the supernova engine. A future detection of neutrinos will allow us to probe mainly the thermodynamic properties of the collapsed core and GWs will allow us to understand the dynamics of moving matter. While neutrinos were already detected from SN~1987A~\cite{hirata:87,bionta:87,1987ESOC...26..237A}, GWs from a CCSN have not yet been observed.

CCSNe are stochastic in nature due to the turbulent flow of matter and so are the predicted detailed GW time series. Detecting these bursts is challenging and requires algorithms that operate when the signals cannot be robustly predicted besides the same general constraints in bandwidth and duration. The detection and reconstruction algorithms should be designed for a large range or even unexpected GW morphologies. So far, few algorithms were used to search for GWs from CCSNe. In 2005, a search targeting CCSN bursts was performed using TAMA300 data using an excess-power filter~\cite{Ando:2004rr}. In 2016, LIGO, Virgo and GEO~600~\cite{Dooley:2015fpa} performed a search~\cite{Abbott:2016tdt} using the excess-power coherent Waveburst (cWB)~\cite{Klimenko2016} and X-pipeline~\cite{Sutton:2009gi} search algorithms. In 2020, LIGO-Virgo conducted an analogues search~\cite{Abbott:2019pxc} using only cWB. The generic all-sky searches also have the potential to detect CCSN GW bursts. Several searches were conducted prior to the observing runs of the advanced detectors, e.g.~\cite{Abbott:2009zi,Abadie:2010mt,Abadie:2012rq}. During O1 and O2, LIGO-Virgo performed searches for GW bursts~\cite{Abbott:2016ezn,Abbott:2019prv} using cWB, oLIB~\cite{Lynch:2015yin}, and BayesWave~\cite{Cornish:2014kda} as a follow-up of the detection candidate events. 

Although the predicted signal morphologies are uncertain, some consensus emerged from the multidimensional supernova simulations~\cite{Mezzacappa:2020oyq,Muller:2020ard,OConnor:2018sti}. This knowledge is useful for improving the existing methods for searches, reconstructing waveforms, and inferring physical properties. Being non-deterministic, matched filtering cannot be used and methods should allow for uncertainties in the signal models. Among these methods are Principal Component Analysis~\cite{Heng:2009zz,Rover:2009ia,2017PhRvD..96l3013P,Roma:2019kcd}, Bayesian inference~\cite{Summerscales:2007xq,Rover:2009ia,2017PhRvD..96l3013P,Roma:2019kcd,Gill:2018hxg}, Machine Learning~\cite{Cavaglia:2020,Portilla:2020gdf,Iess:2020yqj,Chan:2019fuz,Astone:2018uge}, denoising techniques~\cite{Mukherjee:2017}, and others~\cite{Arnaud:1998nk,Hayama:2007iz,Bizouard:2020sws}. These methods apply the knowledge of CCSN models to different degrees. Given the non-deterministic nature of CCSNe and uncertainties of the models, a detection algorithm should use weak or minimal assumptions.

Detecting GWs from exploding stars is a challenge and the search algorithms should be developed before the next nearby CCSN event happens. The cWB algorithm is used regularly in searching for a variety of GWs (e.g.~\cite{Abbott:2019pxc,Abbott:2019prv,Abbott:2016ezn,Salemi:2019owp,Abbott:2019heg}) with minimal assumptions on the signal morphologies. It regularly detects GWs from binary BH  mergers~\cite{LIGOScientific:2018mvr,Abbott:2020niy}, it was the only search algorithm to detect GW150914 in low-latency~\cite{Abbott:2017pqa} and recently it observed the first GW detection of an intermediate-mass binary black-hole GW190521~\cite{Abbott:2020tfl,Szczepanczyk:2020osv}. The cWB search was performing low-latency analysis during each observing run of the advanced detectors and during O3 it was the only algorithm capable of detecting GW bursts in low-latency. It is therefore a promising tool for the first detection of GWs from the next nearby CCSN.

Prospects for the detection of GWs from CCSNe with the Advanced LIGO and Virgo detectors was previously discussed in Gossan et al.~\cite{Gossan:2015xda}. Since then, improvements to the algorithms have resulted in the cWB CCSN search becoming more sensitive~\cite{Abbott:2019pxc} than the previous CCSN analyses performed with the X-pipeline algorithm used in Refs.~\cite{Gossan:2015xda,Abbott:2016tdt}. The detectors were upgraded, for example, light squeezing was introduced~\cite{Tse:2019}, KAGRA joined the network of GW detectors~\cite{Aasi:2013wya}, as well as further upgrades outlined in~\cite{Aasi:2013wya}. The multidimensional CCSN simulations have advanced significantly, including longer duration three-dimensional simulations that predict the entire GW signal, a better coverage of the CCSN parameter space, and an increase in the number of three-dimensional simulations with respect to~\cite{Gossan:2015xda}, and the explosions result in larger GW amplitudes (e.g.~\cite{Radice2019,Andresen_2017}).

Given all these advances, it is important to understand the feasibility of detecting and reconstructing GWs from the next Galactic or near extra-Galactic CCSN with the planned observations and revisit some of the previous results. In this paper, we perform an extensive analysis of simulated state-of-the-art GW signals and make predictions for the fourth and fifth observing runs (O4 and O5). Using a large set of predicted GW signals we provide basic properties, compare their energy evolution, spectra and we list the dominant emission processes. While the previous predictions did not discuss the statistical significance and they relied on data from the initial GW detectors that were available at that time~\cite{Gossan:2015xda}, we use O2 data rescaled to the projected sensitivities of O4 and O5 in such a way that the features of the noise are preserved. Similar to the LIGO-Virgo searches, we perform a background analysis that allows measuring the statistical significance of the detected events. It is important to stress that the statistical significance of a detection statement is fundamental. Any astrophysical evaluation from the reconstructed GW needs to rely on the significance of an event. For the detection sensitivity studies, a fixed significance level allows comparing the performance of different algorithms on the same data set. The results presented in this paper assume that the events are detected at a high significance level.

Given the uncertainties of the predicted GW signals, we use a wide range of models. We significantly expanded the list of analyzed waveforms in comparison to the previous studies of Refs.~\cite{Abbott:2019pxc,Abbott:2016tdt,Gossan:2015xda}, we include signals that became available at the start of the analysis. The adapted waveform families aim to reflect the landscape of GWs and the richness of physical processes in CCSNe. We describe the challenges of detecting these physical processes. We also quantify the reconstruction accuracy for a wide range of GW morphologies.

The paper structure is as follows. The multidimensional CCSN models and the waveforms used in this paper are described in Section~\ref{sec:models}. We highlight the main GW emission processes and the basic properties of the waveforms, such as their duration, energy, or spectrum. Section~\ref{sec:method} outlines the adopted method that consists of the cWB search pipeline, noise rescaling technique, the background estimation, sensitivity analysis, and the procedure to quantify the accuracy of the cWB reconstruction. Section~\ref{sec:results} provides the results. Using data from the LIGO detectors, we determine the distance of a CCSN source and how strong the GW signal should be to be detected by cWB. We specify the reconstruction accuracy of the analyzed waveforms and indicate the challenges regarding the reconstruction of certain GW morphologies. The inclusion of the Virgo and KAGRA~\cite{Aso:2013eba} detectors is also outlined. Finally, Section~\ref{sec:summary} is a summary of the obtained results.


\section{Core-Collapse Supernova}
\label{sec:models}

During its lifetime a massive star burns its fuel by the means of nuclear fusion. A star's structure becomes an onion shape with an iron core in the center. When a core exceeds the Chandrasekhar mass (around 1.4\,$M_\odot$), the gravitational force is so strong that the core collapses, thus forming a very hot proton-neutron star (PNS). Further evolution may lead to an explosion, the collapse to a BH, or a combination of these fates. While the explosion mechanism is currently not settled (see~\cite{Janka:2012wk} for a review), it is believed that the massive flux of neutrinos from the PNS plays a crucial role. In this so-called \textit{neutrino-driven} mechanism, the neutrinos heat up the matter creating a shock that eventually may blow up the star (see~\cite{Mezzacappa:2020oyq,Janka:2017vcp} for a review). When a progenitor star rotates rapidly and has a strong magnetic field, the \textit{magnetorotationally-driven} (MHD-driven) mechanism is more likely to explain a CCSN explosion. During a collapse, a seed magnetic field is largely magnified giving a rise to jets moving along the rotational axis that can contribute to destroying the star (e.g.~\cite{Scheidegger2010,Obergaulinger:2020cqq,Nishimura:2015nca}). In the quantum chromodynamics \textit{phase transition} mechanism, the accreting matter increases PNS density and temperature (e.g.~\cite{Fischer:2020xjl,Zha:2020gjw}). When the PNS collapses, the phase transition to quark matter may occur launching a shock dominating an explosion. In the case when the shock revival mechanism fails or matter continues accreting (\textit{fallback}), a star undergoes \textit{BH formation} (e.g.~\cite{Cerda-Duran2013,Chan2018,Kuroda2018,Pan2018,Pan2020,2021arXiv210106889P,OConnor:2010moj}). In the \textit{extreme emission models} (e.g.~\cite{ottdcc:10,piro:07}), the PNS may be highly deformed due to a very rapid rotation of a progenitor star or even fragmented. While the fraction of CCSNe that form BHs is uncertain, it can be up to 20\%~\cite{OConnor:17,Ugliano:12}. Around 99\% of the explosions are believed to be neutrino-driven and the rest 1\% are MHD-driven if the numbers are correlated with the observed neutron stars and magnetars~\cite{Janka:2012wk,Woosley:2005gy}.

For the neutrino-driven mechanism, the CCSN evolution can be divided into a few phases~\cite{Muller2012,Yakunin2015a,Mezzacappa:2020lsn,Andresen2017}. Here, we describe this evolution using three generic stages. During the first phase, the iron core collapses and bounces when it reaches nuclear densities. The supersonic collapse of the iron core and the infalling external layers launches an initial shock that expands and halts producing a decrease in frequency \textit{prompt-convection} GW signal. This emission is followed by a relatively short quiescent period. The second phase begins around 100\,ms with a strong rise of a neutrino outflow from the hot PNS. The neutrinos deposit energy in the turbulent matter between the PNS and the shock wave pushing it outward. This shock revival mechanism is crucial for a star to explode. For a few hundred milliseconds the effect of the neutrino heating on the shock is competing with the accreting matter. This phase can also induce neutrino-driven convection between the shock and the PNS surface. The shock itself can oscillate (in linear and spiral motion) which is referred to as the standing-accretion shock instability (SASI)~\cite{Blondin:2003}. These aspherical matter movements produce low-frequency \textit{SASI/convection} GW signals in the frequency band where GW detectors are most sensitive. During this phase, the PNS is stiffening over time because of the residual electron capture and it is excited continuously by the accreting matter. The restoring force for these oscillations can be gravity, surface, or pressure (\textit{g-, f-, p-modes}) GW signals. If the shock expansion is fully revived, the explosion phase occurs, and the accretion continues at a smaller rate and the SASI and convection die out.

When a progenitor star rotates rapidly, the explosion mechanism is likely to be MHD-driven. In this scenario, rapid rotation flattens the iron core, producing an axisymmetric collapse and a strong linearly polarized GW \textit{bounce} signal. The latter stages after bounce are not yet well understood because of insufficient MHD microphysics in the numerical simulations (even if very active research is ongoing~\cite{Mezzacappa:2020oyq,Muller:2020ard}). Regardless of the progenitor star rotation, if the shock is not revived and the matter continues to fall, the PNS can collapse further to a BH. The \textit{BH formation} GW signal ends abruptly at the moment of the event horizon creation.


\begin{table*}

\caption{
  Waveforms from multidimensional CCSN simulations described in the text. For each waveform family we provide a reference, dimensionality, a summary of the numerical method (EOS and code name) and observed GW features. Then, we provide details for example waveforms: identifier, progenitor stellar mass $M_\mathrm{star}$, initial central angular velocity $\Omega_c$, the frequency $f_\mathrm{peak}$ at which the GW energy spectrum peaks, the emitted GW energy $E_{GW}$ and approximate signal duration. The superscript symbols: $^\dag$non-ZAMS, $^\star$the simulation was stopped before the full GW signal was developed.}
\begin{tabular}{ccccccccc}
\hline
\hline
\multicolumn{1}{c}{Waveform}
&\multicolumn{1}{c}{Numerical}
&\multicolumn{1}{c}{GW}
&\multicolumn{1}{c}{Waveform}
&\multicolumn{1}{c}{$M_\mathrm{star}$}
&\multicolumn{1}{c}{$\Omega_c$}
&\multicolumn{1}{c}{$f_\mathrm{peak}$}
&\multicolumn{1}{c}{$E_\mathrm{GW}$}
&\multicolumn{1}{c}{Duration}
\\
Family 
& Method
& Features
& Identifier
& [M$_\odot]$
& $[\mathrm{rad}/\mathrm{s}]$
&\multicolumn{1}{c}{[Hz]}
&\multicolumn{1}{c}{[$M_\odot c^2$]}
&\multicolumn{1}{c}{[ms]}\\
\hline
\hline
\multirow{4}{*}{\makecell{Abdikamalov et al.\\ 2014, 2D \cite{Abdikamalov_2014}  }}
& \multirow{4}{*}{\makecell{LS220, Shen\\ CoCoNuT }} 
& \multirow{4}{*}{\makecell{bounce\\ prompt-conv. }}  
& A1O01.0 & $12$ & 1.0 & 819 & \phantom{0}$9.4\times 10^{-9\phantom{0}}$ & $50^*$ \\
& & & A2O01.0 &$12$ & 1.0 & 854 & \phantom{0}$1.7\times 10^{-8\phantom{0}}$ & $50^*$ \\
& & & A3O01.0 &$12$ & 1.0 & 867 & \phantom{0}$7.0\times 10^{-9\phantom{0}}$ & $50^*$\\
& & & A4O01.0 &$12$ & 1.0 & 873 & \phantom{0}$4.2\times 10^{-9\phantom{0}}$  & $50^*$ \\
\hline
\multirow{4}{*}{\makecell{Andresen et al.\\ 2017, 3D \cite{Andresen_2017}}}
& \multirow{4}{*}{\makecell{LS220\\ CoCoNuT\\ PROMETHEUS }}
& \multirow{4}{*}{\makecell{g-modes\\ SASI (spiral)\\ convection}}
&     s11 &$11.2$ & - & 642 & \phantom{0}$1.1\times 10^{-10}$ & $350^*$ \\
& & & s20 &$20$  & - & 687 & \phantom{0}$7.4\times 10^{-10}$ & $430^*$ \\
& & & s20s &$20$ & - & 693 & \phantom{0}$1.4\times 10^{-9\phantom{0}}$ & $530^{*}$ \\
& & & s27  &$27$ & - & 753 & \phantom{0}$4.4\times 10^{-10}$& $570^{*}$   \\
\hline
\multirow{3}{*}{\makecell{Andresen et al.\\ 2019, 3D \cite{Andresen_2019}}}
& \multirow{3}{*}{\makecell{LS220\\ PROMETHEUS }} 
& \multirow{3}{*}{\makecell{SASI (spiral)\\ g-modes}}
& m15fr &$15$ & 0.5 &   689 & \phantom{0}$2.7 \times 10^{-10}$ & $460^{*}$ \\
& & & m15nr &$15$ & - & 820 & \phantom{0}$1.5 \times 10^{-10}$ & $350^{*}$ \\
& & & m15r &$15$ & 0.2 & 801 & \phantom{0}$7.1\times 10^{-11}$ & $380^{*}$ \\
\hline
\multirow{2}{*}{\makecell{Cerd\'a-Dur\'an et al.\\ 2013, 2D \cite{Cerda-Duran2013} }}
& \multirow{2}{*}{\makecell{LS220\\ CoCoNuT }}
& \multirow{2}{*}{\makecell{BH formation\\ g-modes, SASI/conv.}}
& fiducial & 35 & 2.0 & 922 & \phantom{0}$3.3\times 10^{-7\phantom{0}}$ & 1620 \\
& & & slow & 35 & 1.0 & 987 & \phantom{0}$9.4\times 10^{-7\phantom{0}}$ & 1050 \\
\hline
\multirow{3}{*}{\makecell{Dimmelmeier et al.\\ 2008, 2D \cite{Dimmelmeier2008}}}
& \multirow{3}{*}{\makecell{LS, Shen\\ CoCoNuT}}
& \multirow{3}{*}{\makecell{bounce\\ prompt-conv. }}
& s15A2O09-ls & 15 & 4.6 & 743 & \phantom{0}$2.7 \times 10^{-8\phantom{0}}$ & $60^{*}$ \\
& & & s15A3O15-ls & 15 & 13.3& 117 & \phantom{0}$5.2\times 10^{-9\phantom{0}}$ & $340^{*}$ \\
& & & s20A3O09-ls & 20 & 9.0 & 615 & \phantom{0}$2.2\times 10^{-8\phantom{0}}$ & $80^{*}$ \\
\hline
\multirow{2}{*}{\makecell{Kuroda et al.\\ 2016, 3D \cite{Kuroda2016}}}
& \multirow{2}{*}{\makecell{SFHx, DD2, TM1\\ 3D-GR}}
& \multirow{2}{*}{\makecell{g-modes\\ SASI}}
& SFHx & $15$ & - & 718 &  \phantom{0}$2.1 \times 10^{-9\phantom{0}}$ & $350^{*}$  \\
& & & TM1 & $15$ & - & 714 &  \phantom{0}$1.7 \times 10^{-9\phantom{0}}$ & $350^{*}$ \\
\hline
\multirow{2}{*}{\makecell{Kuroda et al.\\ 2017, 3D \cite{Kuroda2017} }}
& \multirow{2}{*}{\makecell{SFHx, DD2, TM1\\ 3D-GR}}
& \multirow{2}{*}{\makecell{g-modes\\ SASI/convection }}
& s11.2 & $11.2$ & - & 195 &  \phantom{0}$1.3 \times 10^{-10}$ & $190^{*}$  \\
& & & s15.0 & $15$ & - & 430 & \phantom{0}$3.1 \times 10^{-9\phantom{0}}$ & $210^{*}$ \\
\hline
\multirow{2}{*}{\makecell{Mezzacappa et al.\\ 2020, 3D \cite{Mezzacappa:2020lsn} }}
& \multirow{2}{*}{\makecell{LS220\\ CHIMERA }}
& \multirow{2}{*}{\makecell{g-, p-modes\\ SASI/convection  }}
& \multirow{2}{*}{\makecell{c15-3D }}
& \multirow{2}{*}{\makecell{15 }}
& \multirow{2}{*}{\makecell{- }}
& \multirow{2}{*}{\makecell{1064 }}
& \multirow{2}{*}{\makecell{\phantom{0}$6.4 \times 10^{-9\phantom{0}}$ }}
& \multirow{2}{*}{\makecell{$420^{*}$ }} \\
&  &   &  & &  &  &  &   \\
\hline
\multirow{4}{*}{\makecell{Morozova et al.\\ 2018, 2D \cite{Morozova2018} }}
& \multirow{4}{*}{\makecell{LS220, DD2, SFHo\\ FORNAX }}
& \multirow{4}{*}{\makecell{f-, g-, p-modes\\ SASI/convection }}
& M10\_LS220 & 10 & - & 1594 & \phantom{0}$2.4 \times 10^{-9\phantom{0}}$ & 1210 \\
& & & M10\_DD2 & 10 & -  & 1544 & \phantom{0}$1.7 \times 10^{-9\phantom{0}}$ & 1700 \\
& & & M13\_SFHo & 13 & - & 976 & \phantom{0}$1.1 \times 10^{-8\phantom{0}}$ & 1360 \\
& & & M19\_SFHo  &  $19$ & - & 1851  & \phantom{0}$6.3 \times 10^{-8\phantom{0}}$ & 1540 \\
\hline
\multirow{3}{*}{\makecell{M\"uller et al.\\ 2012, 3D \cite{Muller2012}}}
& \multirow{3}{*}{\makecell{JM\\ PROMETHEUS}}
& \multirow{3}{*}{\makecell{SASI/convection }}
& L15-3 & 15 & -  & 144 & \phantom{0}$2.2\times 10^{-11}$ & 1400 \\
& & & N20-2 & 20 & -  & 147 & \phantom{0}$1.1\times 10^{-11}$ & 1500 \\
& & & W15-4 & 15 & - &  208 & \phantom{0}$2.5\times 10^{-11}$ & 1300 \\
\hline
\multirow{4}{*}{\makecell{O'Connor\&Couch \\ 2018, 3D \cite{OConnor2018}  }}
& \multirow{4}{*}{\makecell{SFHo\\ FLASH }}
& \multirow{4}{*}{\makecell{g-modes\\ SASI/convection }}
& mesa20 & 20 & - & 1121 & \phantom{0}$6.3 \times 10^{-10}$ & $500^*$ \\
& & & mesa20\_LR & 20 & -  & 1199 & \phantom{0}$2.2 \times 10^{-9\phantom{0}}$ & $650^*$ \\
& & & mesa20\_pert & 20 & - & 1033 & \phantom{0}$9.5 \times 10^{-10}$ & $530^*$ \\
& & & mesa20\_v\_LR & 20 & - & 887 & \phantom{0}$1.0 \times 10^{-10}$ & $480^*$ \\ 
\hline
\multirow{4}{*}{\makecell{Ott et al.\\ 2013, 3D \cite{Ott2013} }}
& \multirow{4}{*}{\makecell{LS220\\ Zelmani }}
& \multirow{4}{*}{\makecell{prompt-conv.\\ g-modes }}
&     s27-fheat1.00 & 27 & - & 836 & \phantom{0}$4.0 \times 10^{-10}$ & 190$^{*}$ \\
& & & s27-fheat1.05 & 27 & - & 385 & \phantom{0}$3.4 \times 10^{-10}$ & 190$^{*}$   \\
& & & s27-fheat1.10 & 27 & - & 340 & \phantom{0}$3.3 \times 10^{-10}$ & 190$^{*}$  \\
& & & s27-fheat1.15 & 27 & - & 839 & \phantom{0}$3.1 \times 10^{-10}$ & 190$^{*}$  \\
\hline
\multirow{2}{*}{\makecell{Powell\&M\"uller\\ 2019, 3D \cite{Powell2019} }}
& \multirow{2}{*}{\makecell{LS220\\ CoCoNuT-FMT}}
& \multirow{2}{*}{\makecell{g-modes }}
& s3.5\_pns & $3.5^\dag$ & - & 878 &  \phantom{0}$3.6 \times 10^{-9\phantom{0}}$ & 700  \\
& & & s18 & $18$ & - & 872 &  \phantom{0}$1.6 \times 10^{-8\phantom{0}}$ & 890 \\
\hline
\multirow{3}{*}{\makecell{Powell\&M\"uller\\ 2020, 3D \cite{Powell2020}}}
& \multirow{3}{*}{\makecell{LS220\\ CoCoNuT-FMT}}
& \multirow{3}{*}{\makecell{f-, g-modes\\ SASI\\ prompt-conv. }}
& s18np & $18$ & 3.4 & 742 &  \phantom{0}$7.7 \times 10^{-8\phantom{0}}$ & 1000  \\
& & & m39 & $39$ & - & 674 &  \phantom{0}$7.5 \times 10^{-10}$ & 560 \\
& & & y20 & $20$ & - & 872 &  \phantom{0}$1.0 \times 10^{-8\phantom{0}}$ & 980 \\
\hline
\multirow{3}{*}{\makecell{Radice et al.\\ 2019, 3D \cite{Radice2019} }}
& \multirow{3}{*}{\makecell{SFHo\\ FORNAX }}
& \multirow{3}{*}{\makecell{f-, g-modes\\ SASI/convection \\ prompt-conv. }}
&      s9 & 9  & - &  727 & \phantom{0}$1.6 \times 10^{-10}$ & 1100 \\
& & & s13 & 13 & - & 1422 & \phantom{0}$5.9 \times 10^{-9\phantom{0}}$ & 800$^*$ \\
& & & s25 & 25 & - & 1132 & \phantom{0}$2.8 \times 10^{-8\phantom{0}}$ & 600$^*$ \\
\hline
\multirow{4}{*}{\makecell{Richers et al.\\ 2017, 2D \cite{Richers2017} }}
& \multirow{4}{*}{\makecell{18 EOSs\\ CoCoNuT }}
& \multirow{4}{*}{\makecell{bounce\\ prompt-conv. }}
&     A467\_w0.50\_SFHx & $12$ & 0.5  & 891 & \phantom{0}$1.6 \times 10^{-8\phantom{0}}$ & 60$^*$ \\
& & & A467\_w0.50\_LS220 & $12$ & 0.5 & 820 & \phantom{0}$5.1 \times 10^{-9\phantom{0}}$ & 60$^*$ \\
& & & A467\_w9.50\_SFHx & $12$ & 9.5  & 448 & \phantom{0}$4.2 \times 10^{-8\phantom{0}}$ & 60$^*$ \\
& & & A467\_w9.50\_LS220 & $12$ & 9.5 & 863 & \phantom{0}$4.1 \times 10^{-8\phantom{0}}$ & 60$^*$ \\
\hline
\multirow{3}{*}{\makecell{Scheidegger et al.\\ 2010, 3D \cite{Scheidegger2010} }}
& \multirow{3}{*}{\makecell{LS180\\ Pen }}
& \multirow{3}{*}{\makecell{bounce\\ prompt-conv. \\ convection}}
&     R1E1CA\_L & $15$ & 0.3 & 1103 &$1.2\times 10^{-10}$ & 90$^*$ \\
& & & R3E1AC\_L & $15$ & 6.3 & 588 & \phantom{0}$2.2\times 10^{-7\phantom{0}}$ & 110$^*$ \\
& & & R4E1FC\_L & $15$ & 9.4 & 683 & \phantom{0}$3.9\times 10^{-7\phantom{0}}$ & 100$^*$ \\
\hline
\multirow{4}{*}{\makecell{Yakunin et al.\\ 2015, 2D \cite{Yakunin2015a} }}
& \multirow{4}{*}{\makecell{LS220\\ CHIMERA }}
& \multirow{4}{*}{\makecell{g-modes\\ SASI/convection \\ prompt-conv.}}
&     B12 & $12$  & -  & 708 & \phantom{0}$3.4 \times 10^{-9\phantom{0}}$ & 1300 \\
& & & B15 & $15$  & -  & 865 & \phantom{0}$7.9 \times 10^{-9\phantom{0}}$ & 1100 \\
& & & B20 & $20$  & -  & 602 & \phantom{0}$4.2 \times 10^{-9\phantom{0}}$ & 900 \\
& & & B25 & $25$  & -  &1022 & \phantom{0}$1.4 \times 10^{-8\phantom{0}}$ & 1140 \\
\hline
\hline
\end{tabular}
\label{tab:models}
\end{table*}

\subsection{CCSN Models}

For more than 50 years various efforts have been made to understand the mechanism of evolution of supernovae~\cite{Colgate:1966}. Despite the progress in theoretical and numerical simulations, the dynamics of CCSN explosions are not yet fully understood as extremely complex physics poses many open questions and challenges. For many years, the calculations were performed in one-dimensional and two-dimensional (2D) simulations. Significant progress has been made in recent years with many full three-dimensional (3D) self-consistent simulations. Despite a large number of 3D CCSN simulations, the number of publicly available GW signals is limited because of the computational cost of extracting them.

Table~\ref{tab:models} summarizes the basic information about numerical methods, types of GW emissions predicted by the simulations, and properties of example waveforms used in this work. We analyzed 82 waveforms from 18 waveform families. We study waveforms from 2D and 3D simulations that were available at the beginning of the analysis. We do not analyze GW signals from older simulations (e.g.~\cite{Marek:2008qi,Murphy:09,Yakunin:2010fn,Rampp:1997em,Zwerger:97}) and those that became available during our analysis (e.g.~\cite{Bugli:2019rax,Obergaulinger:2019iyi,Pan:2017tpk,Pan:2020idl,2021arXiv210106889P}). The set reflects the landscape of available GW signals for a variety of progenitor star parameterizations, physics approximations, and GW signal properties. 

The approximations in the numerical setup of the simulations affect GW production. The axisymmetric 2D models produce by definition linearly polarized signals ($h_+$ and $h_\times=0$), while two polarizations ($h_+$ and $h_\times$) are available for 3D simulations. The equation of state (EOS) of the dense matter is an important ingredient, they can range from softer to stiffer and they may alter GW signatures. The EOSs mentioned in Table~\ref{tab:models} are: LS, LS180, LS220~\cite{Lattimer:1991nc}; Shen~\cite{Shen:11}; DD2, TM1~\cite{Hempel:2009mc}; SFHx, SFHo~\cite{Steiner:2012rk}; and others. Various efforts are conducted for approximating General Relativity, neutrino treatment and other physical processes. Some of the approaches used to calculate waveforms we analyze are: CoCoNuT~\cite{Dimmelmeier:05}, CoCoNuT-FMT~\cite{Mueller:2010nf}, PROMETHEUS~\cite{Fryxell:91}, CHIMERA~\cite{Bruenn:2018wpz}, FLASH~\cite{Dubey:09}, Zelmani~\cite{Ott:2012kr}, JM (Janka \& M\"uller~\cite{1996A&A...306..167J}), Pen (Pen et al.~\cite{Pen:2003up}), and 3D-GR~\cite{Kuroda:2015bta}.

For all waveforms, we provide information about the progenitor star masses $M_\mathrm{star}$ that range from $3.5\,M_\odot$ to $60\,M_\odot$. The $3.5\,M_\odot$ progenitor is an ultra-stripped helium star and all other progenitors have zero-age main-sequence (ZAMS) masses. The rotation of the stars is modeled to be differential and initial central angular velocity $\Omega_c$ is provided. The peak frequency $f_\mathrm{peak}$ is calculated from the energy spectra and the GW energy $E_\mathrm{GW}$ is the source angle averaged. The waveform duration is the time from the moment of the collapse until the end of the simulations. Due to a large computational cost, some of the simulations are stopped before the full GW signal develops. This is marked in the table.

{\it Abdikamalov et al. 2014} \cite{Abdikamalov_2014} (Abd+14) study extensively the influence of the angular momentum distribution on the GW signal of rotating collapse, bounce, and the very early postbounce ring-down phase. We analyzed 6 waveforms: AbdA1O01.0, AbdA2O01.0, AbdA3O01.0, AbdA3O06.0, AbdA4O01.0, AbdA5O01.0. Abd+14 do not investigate the post bounce turbulence and its GW production.

{\it Andresen et al. 2017} \cite{Andresen_2017} (And+17) study GW signals from 3D neutrino hydrodynamics simulations of CCSNe. GW emission in the pre-explosion phase strongly depends on whether the post-shock flow is dominated by SASI and g-mode frequency components in their signals. 

{\it Andresen et al. 2019} \cite{Andresen_2019} (And+19) study the impact of moderate progenitor rotation on the GW signals. The stellar evolution calculations include magnetic fields with low angular momentum. GW emission in the pre-explosion phase strongly depends on whether the post-shock flow is dominated by the SASI with neutrino transport and g-mode frequency components in their signals.

{\it Cerd\'a-Dur\'an et al. 2013} \cite{Cerda-Duran2013} (Cer+13) analyze GW emission of the BH formation in the collapsar scenario. The model consists of a rapidly-rotating progenitor with LS220. GW emission in the pre-explosion phase strongly depends on whether the post-shock flow is dominated by the SASI/convection and g-mode frequency components in their signals.

{\it Dimmelmeier et al. 2008} \cite{Dimmelmeier2008} (Dim+08) conduct extensive studies of rotating core collapse and the impact of the rotational profiles, progenitor masses, and EOS. The GW signal is dominated by the core bounce and prompt-convection, depending primarily on the rotation. We analyzed 6 waveforms: s15a2o05\_ls, s15a2o09\_ls, s15a3o15\_ls, s20a1o05\_ls, s20a3o09\_ls, s20a3o13\_ls. Similarly to Abd+10, the post bounce signal is not investigated in this simulation.

{\it Kuroda et al. 2016} \cite{Kuroda2016} (Kur+16) study the impact of the EOSs on the
GW signatures using a $15 M_\odot$ progenitor star. GW emission in the pre-explosion phase strongly depends on whether the post-shock flow is dominated by the SASI/convection and g-mode components in their signals. For the TM1 waveform, only one angle orientation was available to analyze.

{\it Kuroda et al. 2017} \cite{Kuroda2017} (Kur+17) is a continuation of Kur+16 work. Two additional explosions are analyzed, with 11.2\,$M_{\odot}$ and 40\,$M_{\odot}$ progenitor stars. Their study suggests a correlation between neutrino fluxes and GWs from the SASI. For both waveforms only one angle orientation was available.

{\it Mezzacappa et al. 2020}  \cite{Mezzacappa:2020lsn} (Mez+20) study the details of the GW emission origins and their results replace those of Yakunin et al.~\cite{Yakunin:2017tus}. The GW signals have two key features: low-frequency emission ($<200$\,Hz) that emanates from the gain layer as a result of neutrino-driven convection and the SASI and high-frequency emission ($>600$\,Hz) that emanates from the PNS due to convection within it.

{\it Morozova et al. 2018} \cite{Morozova2018} (Mor+18) explore the impact of progenitor star mass, rotation,  EOS, and neutrino microphysics on the GW signatures. Depending on the setup, they find f-, g- and p-modes. We analyzed 8 waveforms: M10\_LS220, M10\_LS220\_no\_manybody, M10\_SFHo, M10\_DD2, M13\_SFHo, M13\_SFHo\_multipole, M13\_SFHo\_rotating, and M19\_SFHo.

{\it M\"uller et al. 2012} \cite{Muller2012} (Mul+12) study the neutrino and GW signatures from  neutrino-driven explosions. The GW signatures are dominated by the low-frequency (100-500\,Hz) convective matter movement.

{\it O'Connor \& Couch 2018} \cite{OConnor2018} (Oco+18) analyze the impact of the progenitor asphericities, grid resolution and symmetry, dimensionality, and neutrino physics. The GW signals are dominated by the g-mode and the SASI activity is strong. We analyze 7 waveforms: mesa20, mesa20\_LR, mesa20\_pert, mesa20\_pert\_LR, mesa20\_v\_LR, mesa20\_2D, and mesa20\_2D\_pert. 

{\it Ott et al. 2013} \cite{Ott2013} (Ott+13) study the post-core-bounce phase focusing on SASI and neutrino-driven convection development. Shortly after the bounce, the cores are strongly deformed by the prompt-convection that dominates the GW emission.

{\it Powell \& M\"uller 2019} \cite{Powell2019} (Pow+19) analyze models with low and regular CCSN explosion energies, and perform simulations covering all evolution phases. Both GW signals show emissions from g-modes that peak at high frequencies. 

{\it Powell \& M\"uller 2020} \cite{Powell2020} (Pow+20) study explosion properties of three progenitor star masses including the impact of rotation in the m39 model. The waveforms from the m39 and y20 models produce very strong GW emissions due to the rapid rotation and very strong neutrino-driven convection, respectively. The s18np model is the same as the s18 model in Pow+19, but without perturbations, which prevents shock revival and produces strong SASI.

{\it Radice et al. 2019} \cite{Radice2019} (Rad+19) explore the dependence of the GW properties on the  progenitor star mass, which ranges from 9\,$M_\odot$ to 60\,$M_\odot$. The signals are dominated by f- and g-modes, but some of them also show strong SASI or prompt-convection signatures. We analyzed 10 waveforms: s9, s10, s11, s12, s13, s14, s15, s19, s25, s60.

{\it Richers et al. 2017} \cite{Richers2017} (Ric+17) perform an extensive analysis of the bounced signal. They show that the signal is largely independent of the choice of EOS, but it is sensitive to the rotational parameters. We analyzed 12 waveforms: A467w0.50\_BHBLP, A467w0.50\_GShenFSU2.1, A467w0.50\_HSDD2, A467w0.50\_LS220, A467w0.50\_SFHo, A467w0.50\_SFHx, A467w9.50\_BHBLP, A467w9.50\_GShenFSU2.1, A467w9.50\_HSDD2, A467w9.50\_LS220, A467w9.50\_SFHo, A467w9.50\_SFHx. As for Abd+14 and Dim+08, the post-bounce phase is not simulated.

{\it Scheidegger et al. 2010} \cite{Scheidegger2010} (Sch+10) show a systematic study of GW signatures from MHD-driven explosions. They study the effects of the EOS, initial rotational rate, and the magnetic field. We analyze three waveforms that vary in rotation.

{\it Yakunin et al. 2015} \cite{Yakunin2015a} (Yak+15) study the full GW evolution from simulations with four progenitor star masses. These waveforms capture several stages of the explosion. All GW signals show both low (SASI/convection) and high (g-mode) frequency components.

\begin{figure*}[ht] 
\includegraphics[width=0.70\linewidth]{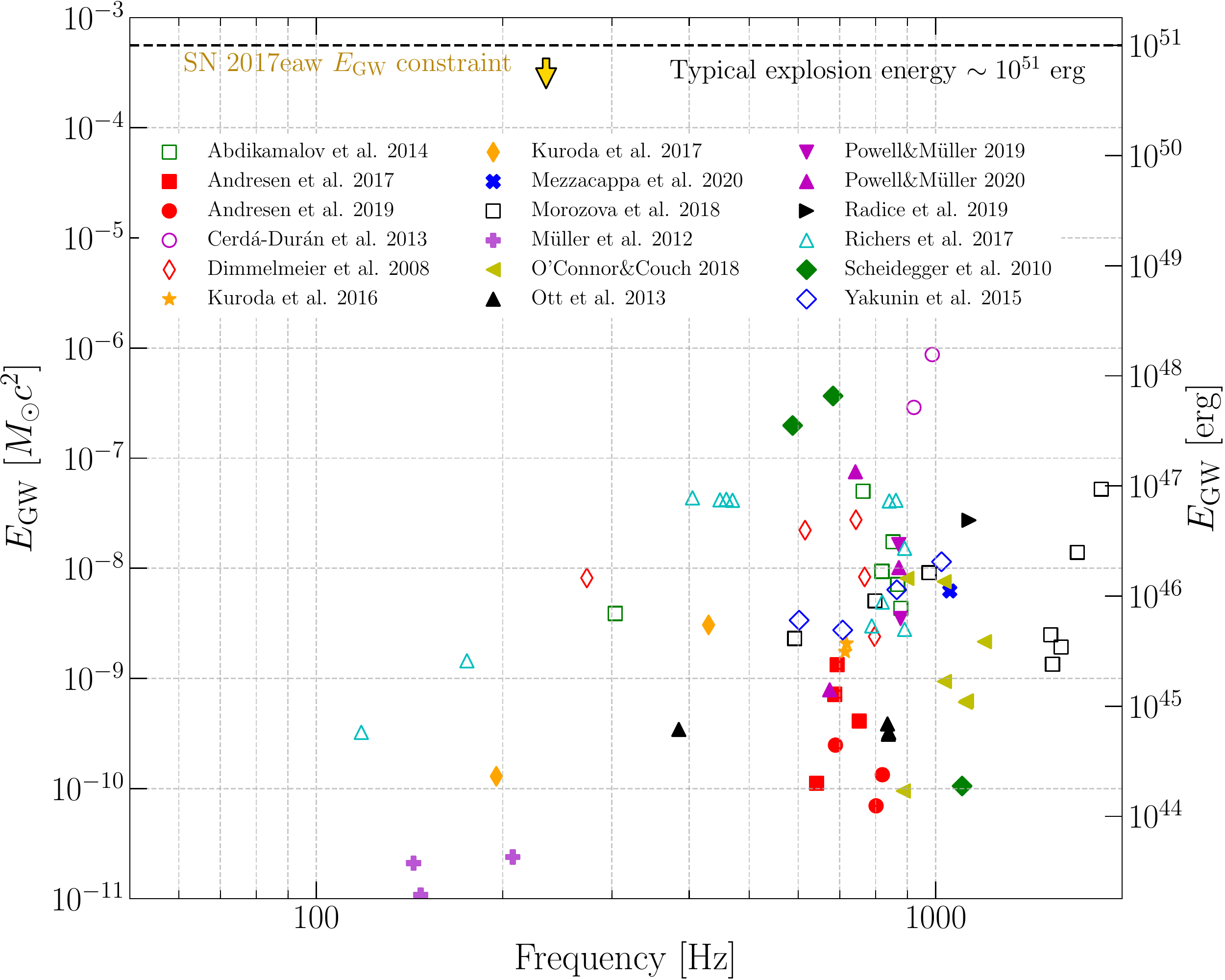}
\caption{GW energy as a function of peak frequency (maximum of $dE_\mathrm{GW}/df$) for 82 analyzed waveforms. The signals from 2D models are shown with hollow symbols. Spectra of some waveforms are wide-band and the peak frequencies could not be accurately determined. For the majority of the signals, the peak frequencies lay between 300\,Hz and 1000\,Hz that usually corresponds to the proto-neutron star oscillations. The typical energy is in the range from $10^{-10}$ to $10^{-7}M_\odot c^2$ that is smaller than 0.01\%  of a typical CCSN explosion energy. The current GW energy constraint is $4.27\times10^{-4}M_\odot c^2$ at 235\,Hz~\cite{Abbott:2019pxc}. }
\label{fig:egw}
\end{figure*}

\subsection{Gravitational wave calculations}

The quadrupole approximation is commonly used to extract GWs generated by the accelerating matter in CCSNe. The quadrupole radiation is extensively described in the literature (e.g.~\cite{Thorne:80,Maggiore:1900zz,Oohara:1997}). The metric perturbation $\mathbf{h}^{TT}_{ij}$ in the transverse-traceless ($TT$) gauge can be expressed as:
\begin{equation}
    \mathbf{h}^{TT}_{ij}(t,\mathbf{x}) = \frac{1}{D} \ddot{Q}_{ij}(t-D/c,\mathbf{x}),
    \label{eq:hij}
\end{equation}
where $i,j=\{1,2,3\}$ are indices in Cartesian coordinates, $c$ is the speed of light, $D$ is the distance to the source and the dots represent the second time derivative. The traceless quadrupole moment $Q_{ij}$ is defined as:
\begin{equation}
    Q_{ij}(t,\mathbf{x}) = \frac{2G}{c^4} \int d^3x \rho(t,\mathbf{x}) ( x_i x_j - \frac{1}{3} \delta_{ij} |\mathbf{x}|^2 ),
\end{equation}
where $\rho$ is a mass density, $\delta_{ij}$ Dirac delta and $G$ the gravitational constant. For practical reasons, usually the traceless $\ddot{Q}_{ij}$ is directly extracted from the  multidimensional CCSN simulations. However, it is not always the case and in our analysis we unified all the outputs from the CCSN simulations into traceless $\ddot{Q}_{ij}$.

The metric perturbation can also be written as:
\begin{equation}
    \mathbf{h}^{TT}_{ij} = h_+ \mathbf{e}_+ + h_\times \mathbf{e}_\times,
\end{equation}
where $\mathbf{e}_+$ and $\mathbf{e}_\times$ are unit plus and cross polarization tensors. Using a coordinate transformation between Cartesian and spherical coordinates, the GWs radiated in the $(\theta,\phi)$ direction are expressed as~\cite{Oohara:1997}:
\begin{eqnarray}
    h_{+}      &=& \ddot{Q}_{\theta \theta}-\ddot{Q}_{\phi \phi}, \\
    h_{\times} &=& 2 \ddot{Q}_{\theta \phi}, 
    \label{eq:strain}
\end{eqnarray}
where:
\begin{eqnarray}
\label{eq:qtp}
\ddot{Q}_{\theta \phi} &=&  \left (\ddot{Q}_{22} - \ddot{Q}_{11} \right ) \cos{\theta}\sin{\phi}\cos{\phi} \nonumber \\
& & + \ddot{Q}_{12} \cos{\theta} \left (\cos^2 \phi - \sin^2 \phi \right ) \nonumber \\ 
& & + \ddot{Q}_{13} \sin \theta \sin \phi - \ddot{Q}_{23} \sin \theta \cos\phi,
\end{eqnarray}
\begin{eqnarray}
\ddot{Q}_{\phi \phi} &= \ddot{Q}_{11} \sin^2 \phi + \ddot{Q}_{22} \cos^2 \phi - 2 \ddot{Q}_{12} \sin{\phi}\cos{\phi},
\end{eqnarray}
and
\begin{eqnarray}
\ddot{Q}_{\theta \theta} &= \left ( \ddot{Q}_{11} \cos^2 \phi + \ddot{Q}_{22} \sin^2 \phi +  2 \ddot{Q}_{12} \sin{\phi} \cos{\phi} \right) \cos^2 \theta \nonumber \\
&+ \ddot{Q}_{33} \sin^2 \theta - 2 \left (\ddot{Q}_{13} \cos{\phi} + \ddot{Q}_{23} \sin{\phi} \right ) \sin{\theta} \cos{\theta}. \nonumber \\
\end{eqnarray}

In the case of axisymmetric 2D simulations, the cross polarization is zero. The $Q_{ij}$ matrix has only diagonal components, $Q_{11} = Q_{22} = -\frac{1}{2} Q_{33}$, and the GW strain $h_+$ is related to $\ddot{Q}_{ij}$ as~\cite{Finn:1990}:
\begin{equation}
    h_{+} = \frac{3 \sin^2{\theta}}{2}  \ddot{Q}_{33}.
\end{equation}
where $\theta$ is an inclination angle.

We use $\ddot{Q}_{ij}$ to analyze the waveforms and provide basic properties, such as the total energy, energy evolution, energy spectrum, and the characteristic strain. The total energy is calculated as:
\begin{equation}
    E_{GW}=\int_{-\infty}^{\infty}{\frac{dE_\mathrm{GW}}{dt}}dt
\end{equation}
where~\cite{Maggiore:1900zz,szczepanczykdcc:15}:
\begin{equation}
\begin{split}
    \frac{dE_\mathrm{GW}}{dt}= \frac{G}{5 c^5} ( & \dddot{Q}_{11}^2 
    + \dddot{Q}_{22}^2 + \dddot{Q}_{33}^2 +  \\
        &  2 (\dddot{Q}_{12}^2+ \dddot{Q}_{13}^2 + \dddot{Q}_{23}^2  ) ).
\end{split}
\end{equation}
The energy spectrum is: 
\begin{eqnarray}
    \frac{dE_\mathrm{GW}}{df}= \frac{G}{5 c^5} & (2 \pi f)^2 ( |\ddot{\tilde{Q}}_{11}|^2 + |\ddot{\tilde{Q}}_{22}|^2 + |\ddot{\tilde{Q}}_{33}|^2 + \nonumber \\ 
    & 2 (| \ddot{\tilde{Q}}_{12}|^2 + | \ddot{\tilde{Q}}_{13} |^2 + | \ddot{\tilde{Q}}_{23} |^2 ) ), 
\end{eqnarray}
where $\tilde{Q}_{ij}$ is a Fourier transform of $Q_{ij}$:
\begin{equation}
    \tilde{Q}_{ij}(f) = \int Q_{ij}(t) e^{-i 2\pi t f}.
\end{equation}
The characteristic strain is defined as~\cite{Flanagan:1997sx}:
\begin{equation}
    h_{char} = \frac{1}{D} \sqrt{ \frac{2 G}{\pi^2 c^3} \frac{dE_\mathrm{GW}}{df} }.
\end{equation}

Before the waveforms can be analyzed with cWB, they need to be prepared to avoid analysis artifacts. The CCSN simulations often create GW components below around 10\,Hz that, because of the truncation of the simulations, produce discontinuities at the end of the waveform. This effect is observed to be significant for waveforms from Mor+18, Mul+12, Yak+15 and Pow+20. This low-frequency component is removed here using a high-pass filter with a cutoff of 10\,Hz. All waveforms are resampled to a sampling frequency of 16384\,Hz and they are rescaled to a source distance of 10\,kpc. For 3D simulations, 100 signals are calculated depending on the source orientation, while 10 inclination angles are chosen for 2D simulations. We note that some of the theoretical properties given in this paper differ from those presented in the papers of the corresponding waveforms. It might be caused by the different processing method. However, these discrepancies have an insignificant effect on our results and conclusions.

\begin{figure}[hbt] 
\includegraphics[width=1.0\linewidth]{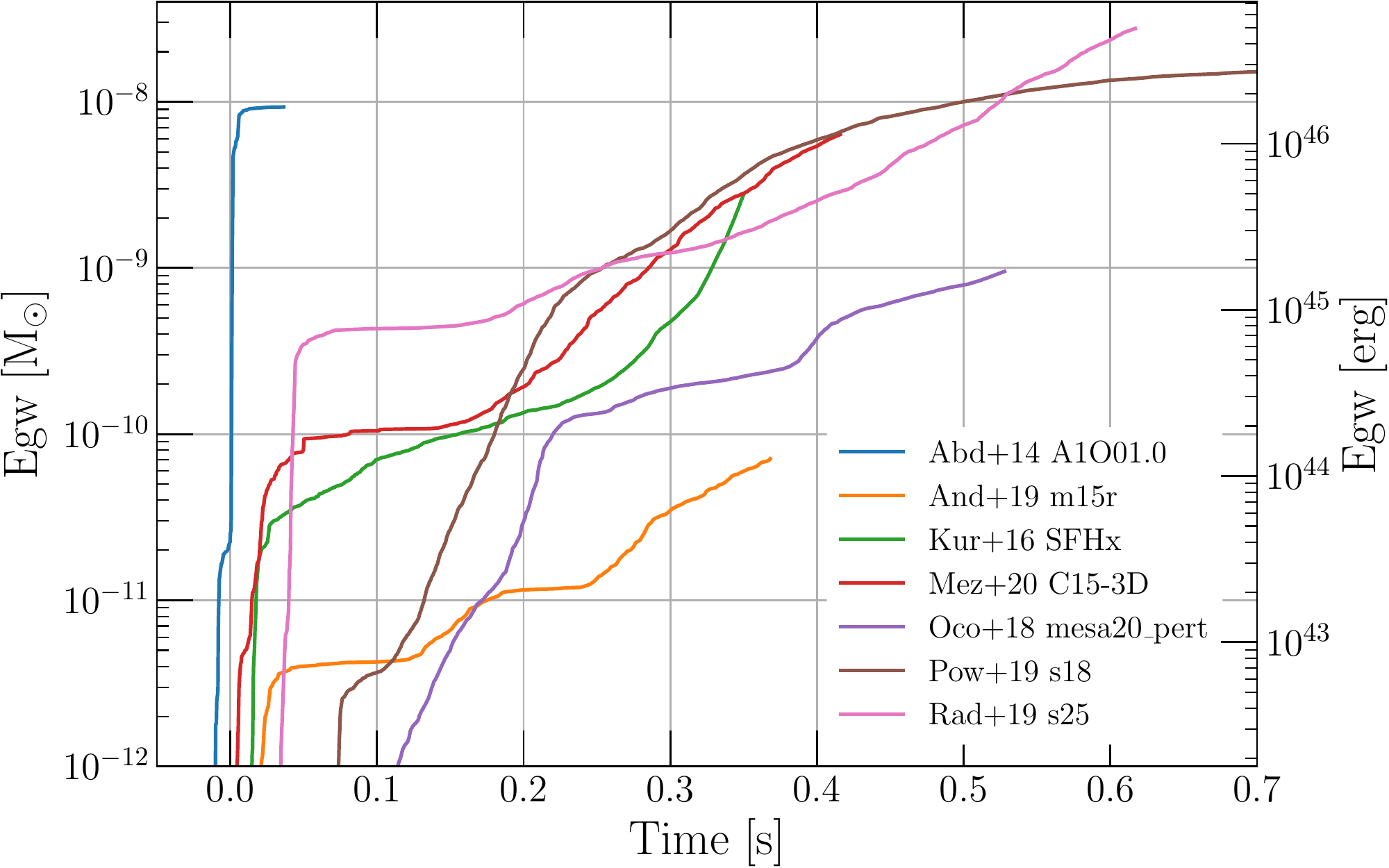}
\caption{Examples of the GW energy evolution. The Abd+14 waveforms are short and energetic core-bounce GW signals. For the neutrino-driven explosions most of the energy is emitted after around 100\,ms. }
\label{fig:egw_evol}
\end{figure}

\subsection{Energy and spectra of GW signals}

The binary BHs are very efficient GW sources, for example, GW150914 radiated around 3\,$\mathrm{M}_\odot c^2$ of energy during the merger~\cite{Abbott:2016blz}. On the other side, GW energies from CCSNe are orders of magnitude weaker. Energies of an order of even $10^{-3}\mathrm{M}_\odot c^2$ can be generated by supernova cores in extreme cases due to the rapid rotation or core fragmentation~\cite{ottdcc:10,piro:07}. However, the energies are significantly smaller for the neutrino- and MHD-driven explosions or BH formations (even if a small fraction of the waveforms is currently available for MHD simulations).

Figure~\ref{fig:egw} shows the source orientation averaged GW energy as a function of the peak frequency $f_\mathrm{peak}$ (frequency of the $dE_\mathrm{GW}(f) / df$ maximum value) for all analyzed waveforms. In the plot, we show the typical explosion energy of a CCSN that is $10^{51}$\,erg (approximately kinetic energy of the ejecta), and the current best GW energy constraint at low frequency is below this limit ($4.27\times10^{-4}M_\odot c^2$ at 235\,Hz~\cite{Abbott:2019pxc}). Energies of most of the waveforms are in the $10^{-10}-10^{-7}\mathrm{M}_\odot c^2$ range with more energetic emissions involving rapid rotation (Cer+13, Sch+10 and Ric+17). Only less than 0.01\% of the explosion energy appears to be converted into GWs. The peak frequencies range from 100\,Hz to above 2\,kHz with the majority of the energy emitted around 1\,kHz.

\begin{figure}[hbt] 
\includegraphics[width=1.0\linewidth]{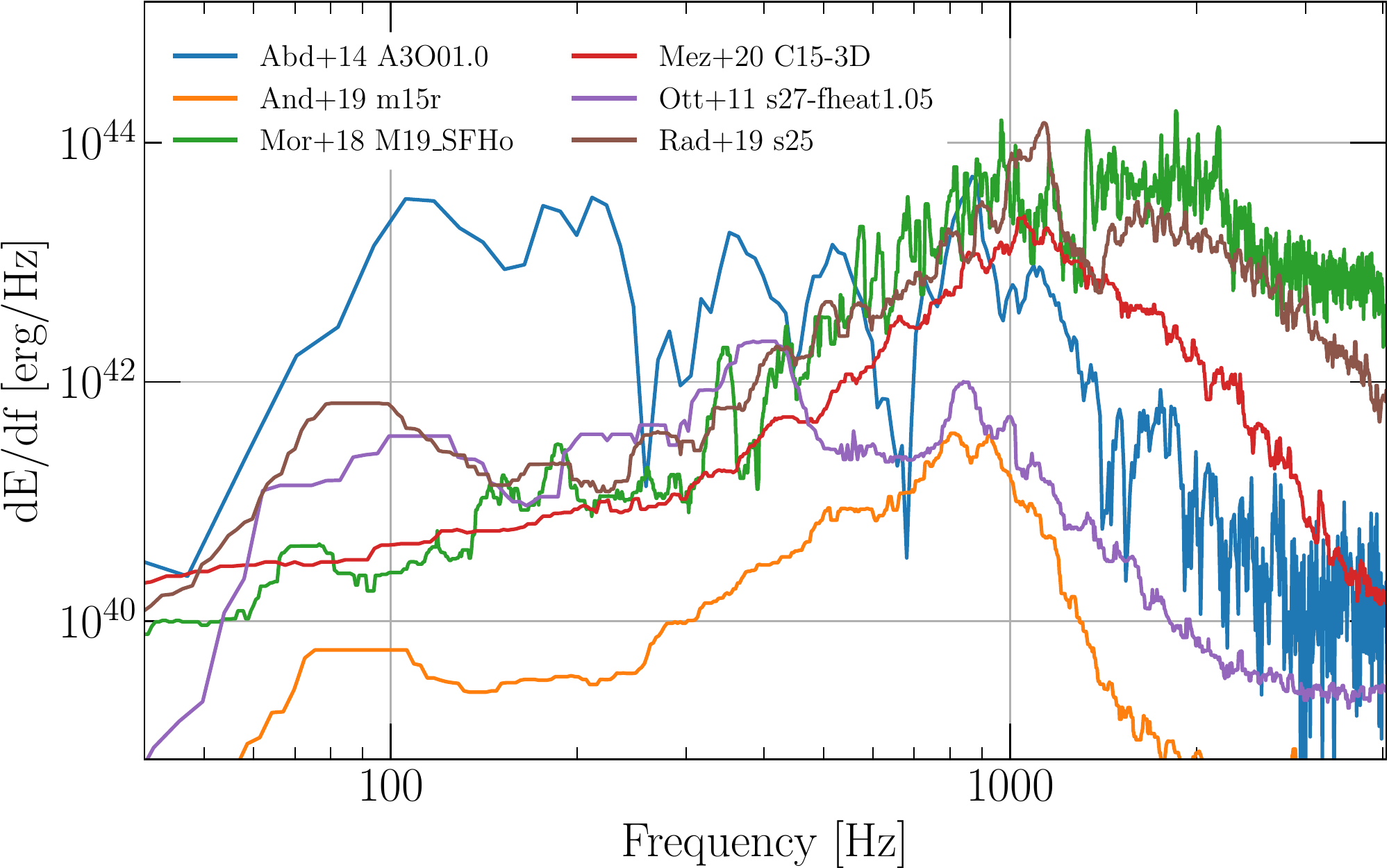}
\caption{The GW signals from CCSNe are typically broadband with the majority of the energy at higher frequencies. The peak frequency can be difficult to determine for some waveforms, like for Abd+14 A3O01.0.}
\label{fig:dedf}
\end{figure}

\begin{figure*}[ht] 
\includegraphics[width=0.70\linewidth]{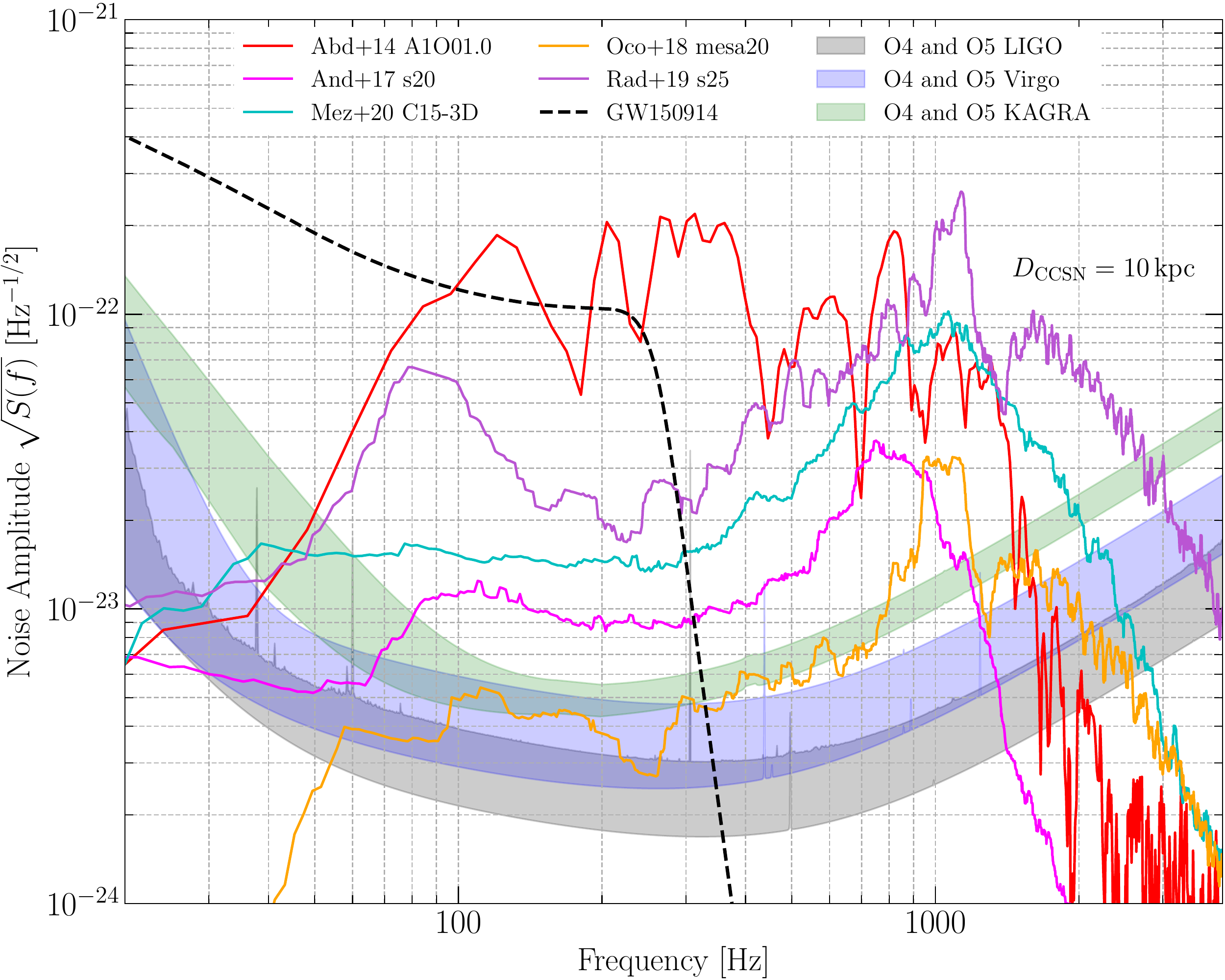}
\caption{The projected noise amplitudes of the LIGO, Virgo, and KAGRA detectors for O4 and O5. The characteristic strains are broadband with the majority of the energy in higher frequencies that usually is emitted by PNS oscillations. Abd+14 represents the core-bounce signal that is strong but broadband. The GW150914 signal is shown for comparison.}
\label{fig:hchar}
\end{figure*}

Figure~\ref{fig:egw_evol} shows example curves of the cumulative energy emitted in GWs as a function of time after core bounce. As described earlier, a CCSN explosion can be divided into a few phases that can be observed in the curves. A core bounce and quiescent phase are followed by a period of accretion and strong GW emission until an explosion phase occurs with typically little accretion and weak GWs. The timescales and the strengths differ between waveforms. Since many simulations are stopped abruptly due to the high computational cost, the GW evolution is stopped before the full signal is evolved. For example, the Abd+14 waveforms represent the bounce signal of a rapidly rotating core and the later evolution is not yet well understood.

Figure~\ref{fig:dedf} shows the GW energy spectra $dE_\mathrm{GW}/df$ for a few example waveforms. The GW signals are usually broadband with the majority of the energy at higher frequencies. The dominant GW emission comes typically from the PNS oscillations. In the case of the Ott+03 model, the explosion is initially very aspherical and the prompt-convection signal around 400\,Hz dominates. In some cases, the peak frequencies cannot be determined unambiguously, for example, the Abd+14 waveforms have multiple peaks in their spectrum.

Figure~\ref{fig:hchar} presents the characteristic strains for example waveforms together with the noise amplitudes of LIGO, Virgo, and KAGRA detectors projected for O4 and O5~\cite{T2000012}. The GW150914 signal is also shown for comparison. The GW detector sensitivities are frequency-dependent and it impacts the detectability of GW features. The stronger GW emission from PNS oscillations peaks in a less sensitive area of the detector spectrum. The GWs from lower frequency SASI/convection have the majority of their energy in the frequency range where the detectors are most sensitive.


\section{Methodology}
\label{sec:method}

\subsection{Coherent WaveBurst}
\label{sec:cwb}

Coherent WaveBurst is an excess-power search algorithm for detecting and reconstructing GWs~\cite{Klimenko2016} using minimal assumptions on the signal morphologies. The cWB search performs a wavelet analysis of GW strain data~\cite{Necula2012}. It selects wavelets with amplitudes above the fluctuations of the detector noise, groups them into clusters, and identifies coherent events.

The cWB event ranking statistics $\eta_c$ is based on the coherent network energy $E_{\rm c}$ obtained by cross-correlating detectors data and $\eta_c$ is approximately the coherent network signal-to-noise ratio. The events are ranked with $\eta_{\rm c}=\sqrt{E_{\rm c}/\max(\chi^2,1)}$. The value of $\chi^2$ quantifies the agreement of cWB reconstruction and the detector data. It is defined as $\chi^2 = E_{\rm n}/N_{\rm df}$, where $E_{\rm n}$ is the residual energy and $N_{\rm df}$ is the number of independent wavelet coefficients used in the reconstruction. $E_{\rm n}$ is the leftover energy after the reconstructed waveform is subtracted. The events are rejected when $\chi^2>2.5$. The further reduction of false alarms due to the non-stationary detector noise is performed using a correlation coefficient $c_{\rm c} = E_{\rm c}/(E_{\rm c}+E_{\rm n})$. The events are accepted when $c_{\rm c}>0.5$. We set up the internal cWB parameters and selection cuts as in the O1-O2 LIGO-Virgo targeted search for CCSNe~\cite{Abbott:2019pxc} with an exception of lowering the $c_{\rm c}$ threshold from 0.8.

\subsection{Noise rescaling}
\label{sec:recolor}

The GW detectors are impacted by many sources of noise. The data is non-stationary, the amplitudes may fluctuate vastly, and it is corrupted by non-Gaussian noise. Every upgrade of the GW interferometers alters the noise properties.  The astrophysical predictions with the projected detector sensitivities should take into account the features of the real detector noise. Therefore, we rescale publicly available O2 data from LIGO Livingston (L1), LIGO Hanford (H1), and Virgo (V1) detectors to the projected sensitivities in O4 and O5~\cite{T2000012}. The data from the KAGRA (K1) detector is not yet available so a Gaussian noise is scaled to projected O4 and O5 sensitivity.

\begin{figure}[hbt] 
\includegraphics[width=0.95\linewidth]{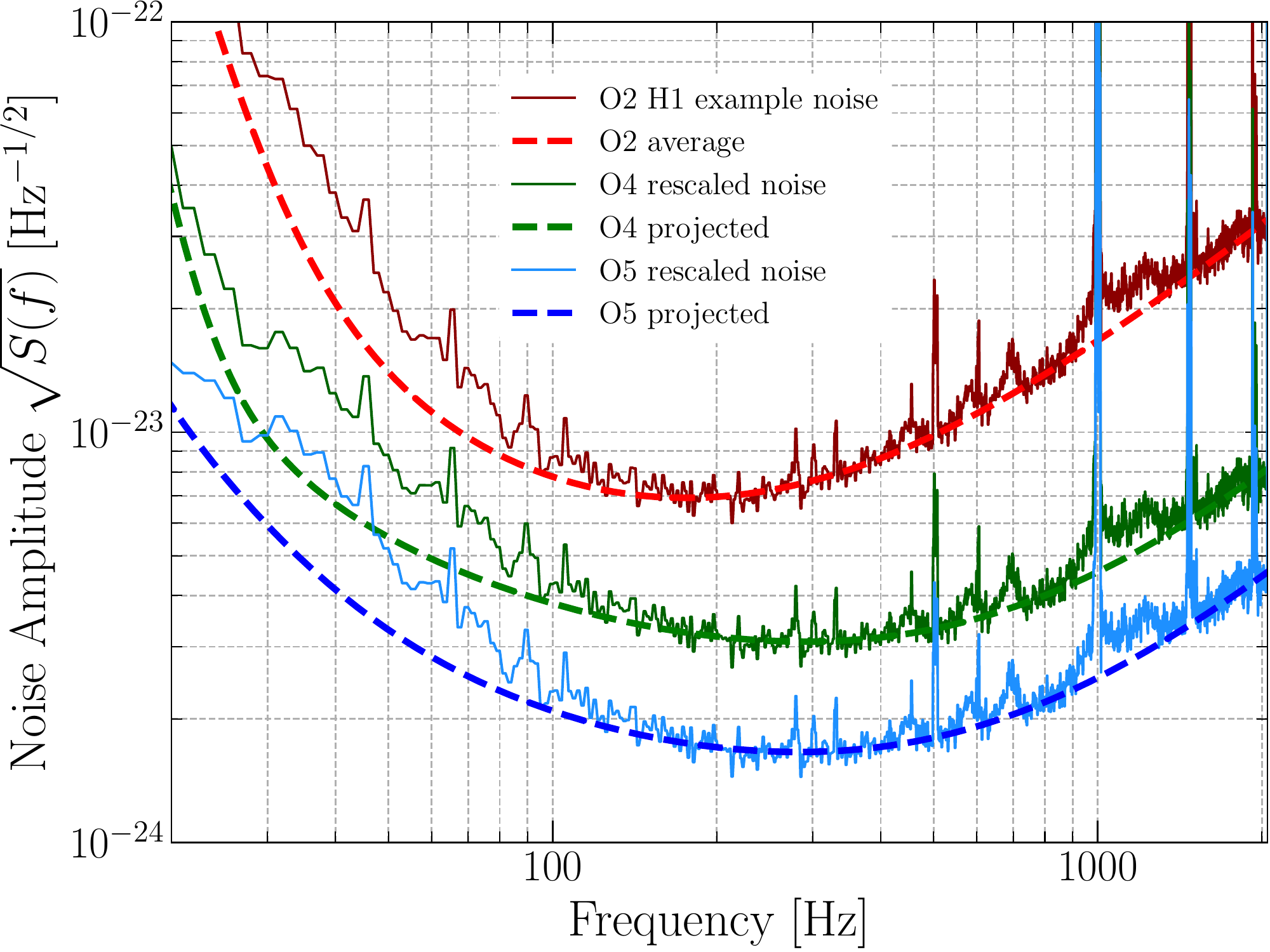}
\caption{Example O2 LIGO Hanford detector sensitivity rescaled to O4 and O5 designs. The rescaling procedure preserves all features of the real noise.}
\label{fig:recolor}
\end{figure}

We developed a procedure that allows us to preserve all features of the noise, including the distributions of glitches, fluctuations of the detector spectra, and other noise sources present in the real data. The rescaling procedure uses an average detector noise spectrum from O2, $S_\mathrm{O2,avg}(f)$~\cite{T1500293,G1801950} and the projected detector sensitivity O5 $S_\mathrm{O5,proj}(f)$. The algorithm takes time series from O2, calculates the spectrum $S_\mathrm{O2}(f)$ and rescales it in the frequency domain as:
\begin{equation}
  S_\mathrm{O5}(f) = S_\mathrm{O2}(f) \frac{S_\mathrm{O5,proj}(f)}{S_\mathrm{O2,avg}(f)}.
\end{equation}
The phase is preserved and the rescaled spectra are transformed back to the time domain. The same procedure is performed with O4 data. Figure~\ref{fig:recolor} shows an example of the spectra of the original H1 O2 data that is rescaled to O4 and O5, an average noise in O2, and the projected O4 and O5 sensitivities. In this example, the algorithm preserves the lower H1 sensitivity below 100\,Hz and a noise excess around 1\,kHz.

\subsection{Background Estimation}

To estimate the background of the noise events we use the same procedure as in~\cite{Abbott:2019pxc}. We perform a time-shifting analysis that is widely used in searches for GW bursts. The data from one detector is shifted a multiple of 1\,s with respect to the other that is longer than the time delay for GW to pass between detectors. It assures that any event identified by cWB is a noise event. The collection of falsely identified events is used to calculate the False Alarm Rate (FAR).

To perform background analysis, we use only H1 and L1 data. The V1 and K1 detectors are predicted to be less sensitive bringing the noise to the coherent analysis. The H1 and L1 detector network (HL) is typically used for background estimation and detection statements in the searches for GW bursts, even if V1 data is available~\cite{Abbott:2019prv,Abbott:2020niy}. The less sensitive detectors may contribute to localizing the GW sources in the sky, but these considerations are beyond the scope of this paper and most likely an exact sky location will be known~\cite{Adams:2013ana}. To estimate the background, we use publicly available O2 LIGO data from the GPS time range corresponding to the search period of SN~2017eaw in~\cite{Abbott:2019pxc}.

\begin{figure}[hbt] 
\includegraphics[width=0.99\linewidth]{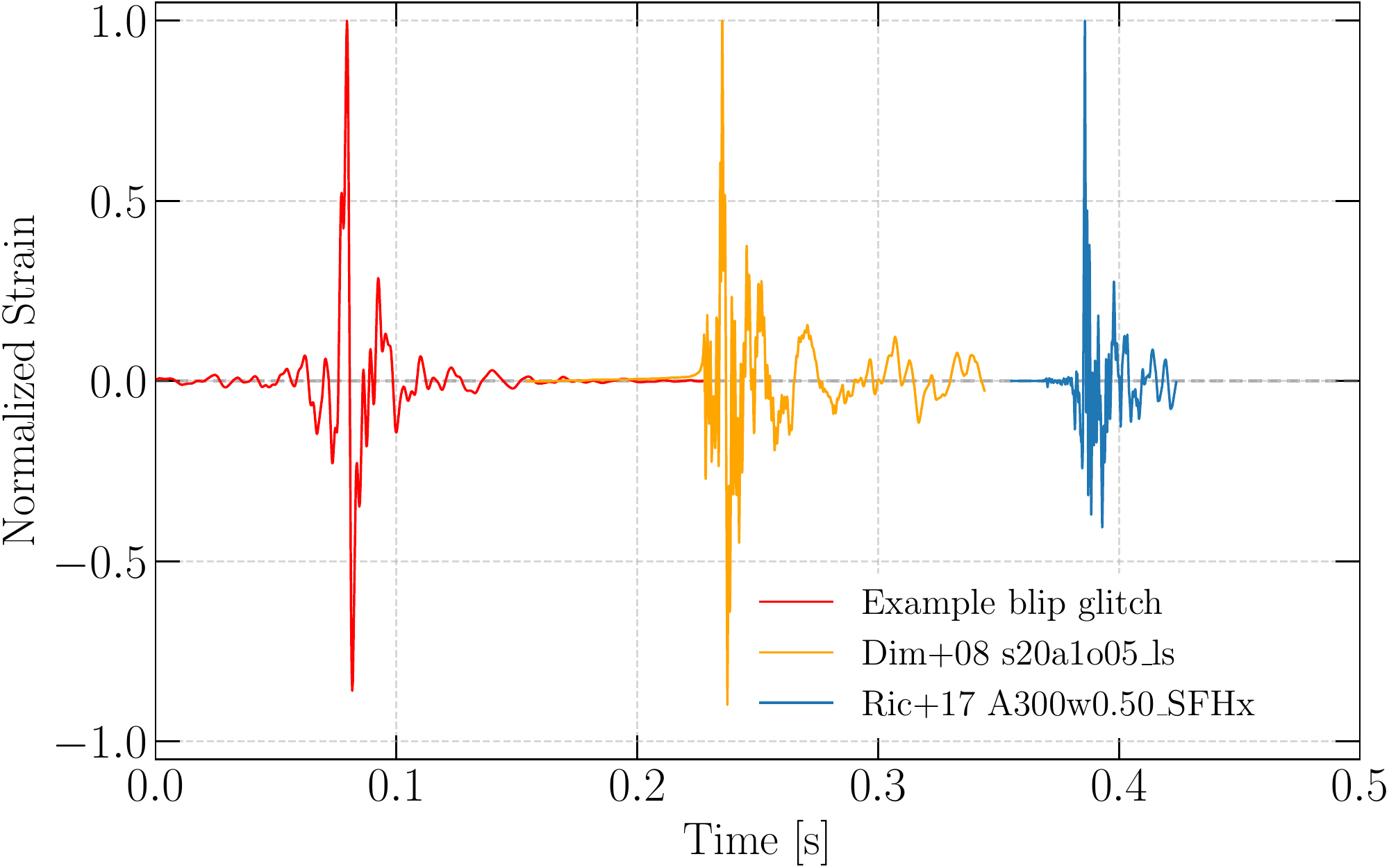}
\caption{The comparison between an example blip glitch and core-bounce waveforms for Dim+08 and Rich+17.}
\label{fig:blip}
\end{figure}

Similar to the search procedure in~\cite{Abbott:2019pxc}, the background events are divided into two mutually exclusive classes. The first class~\textit{C1} contains short-duration transients with up to few cycles, primarily blip glitches~\cite{TheLIGOScientific:2016zmo,Cabero:2019orq} and the second class~\textit{C2} includes other noise events. The blips are $O(10)$\,ms long transients spanning $O(100)$\,Hz frequency band with unknown origin. They are present in all observing runs of the advanced detectors. Importantly, these glitches are morphologically similar to the CCSN bounce signals. Figure~\ref{fig:blip} shows examples of a blip together with Dim+08 and Ric+18 waveforms.

\subsection{Sensitivity studies}

We determine how sensitive the cWB search is to detect and reconstruct CCSN waveforms. The waveforms from different source angle orientations are placed randomly in the sky, added (injected) to the detector noise every 150\,s and reconstructed with cWB. This procedure is performed for a range of source distances creating detection efficiency curves. For each waveform, the distance at 50\% detection efficiency is referred to as a \textit{detection range}. A similar procedure is performed with detection efficiency curves as a function of network signal-to-noise ratio (SNR). The waveforms are placed randomly in the sky and their amplitudes are rescaled to match certain injected SNR (SNR$_\mathrm{inj}$). This allows us to determine how strong the GW signal needs to be to be detected by cWB. The \textit{minimum detectable SNR} is referred to as the SNR at 50\% detection efficiency.

In this search sensitivity study, we use 10~days of coincident data from O2 rescaled to projected O4 and O5 sensitivities. This extended period of data allows us to average the impact of the detector network angular sensitivity and the effects of the noise. We discard events with FAR larger than 1 per year. For a GW signal from a nearby CCSN, the SNEWS~\cite{Antonioli:04} should provide a conservative period of 10\,s to identify the GW burst. Assuming that the GW is detected with FAR smaller than 1 per year, it results in a $5\sigma$ detection confidence (see Eqn.~(1) in~\cite{Abbott:2019pxc}).

\begin{figure*}[hbt]
\begin{minipage}{.5\linewidth}
    \centering
    \includegraphics[width=0.96\linewidth]{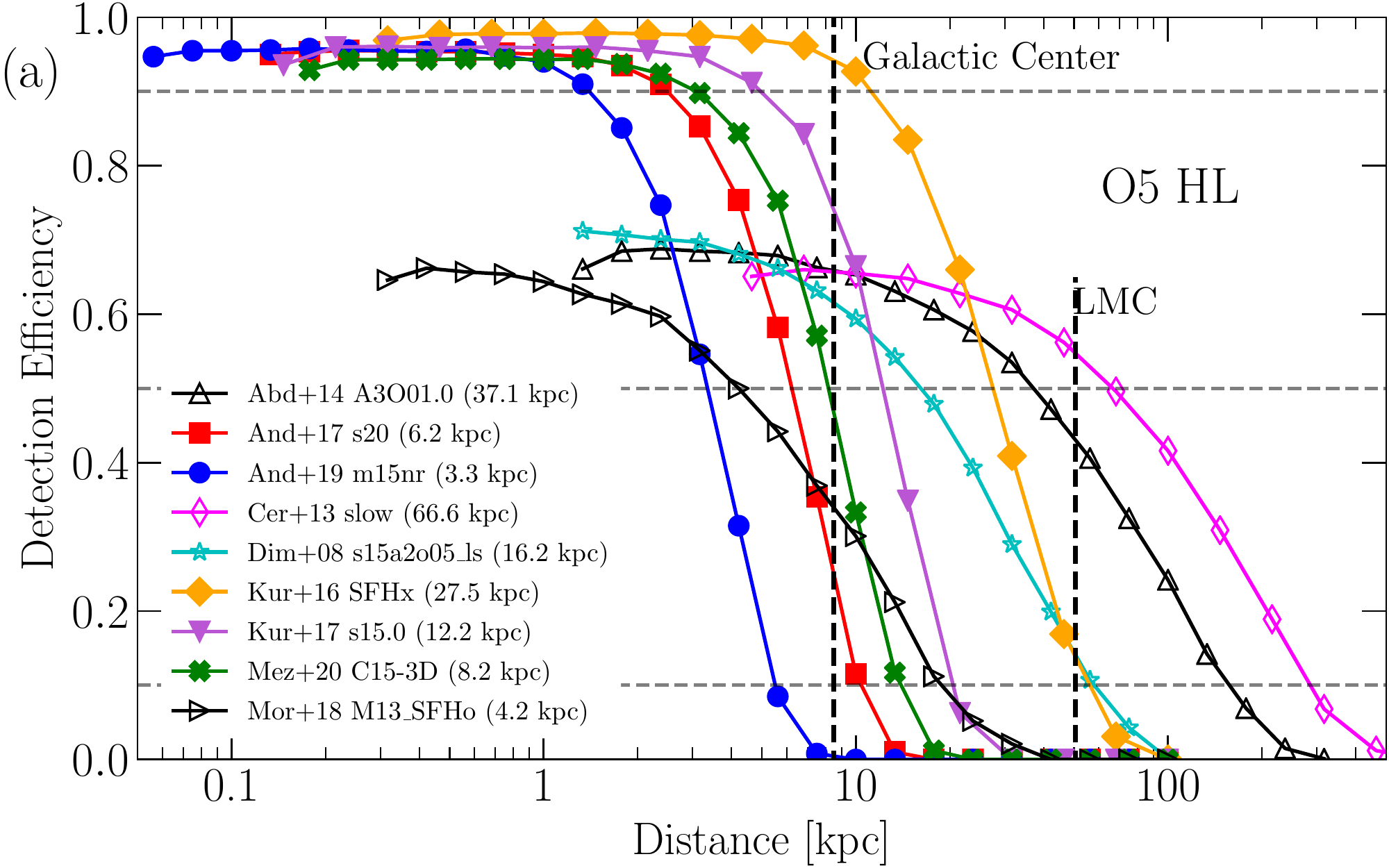}
\label{fig:eff_dist_a}
\end{minipage}%
\begin{minipage}{.5\linewidth}
    \centering
    \includegraphics[width=0.96\linewidth]{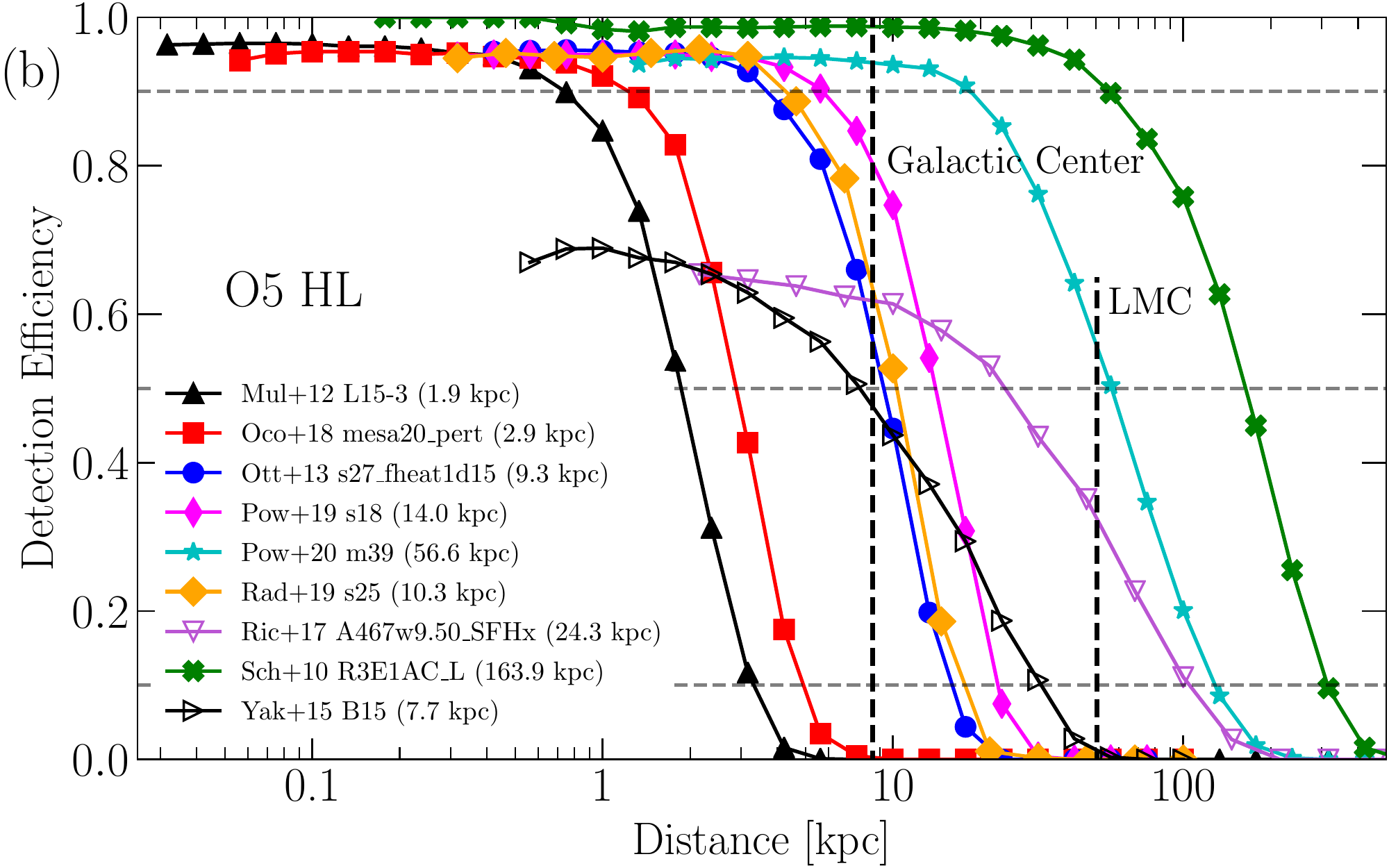}
\end{minipage}\par\medskip
\centering
\includegraphics[width=0.6\linewidth]{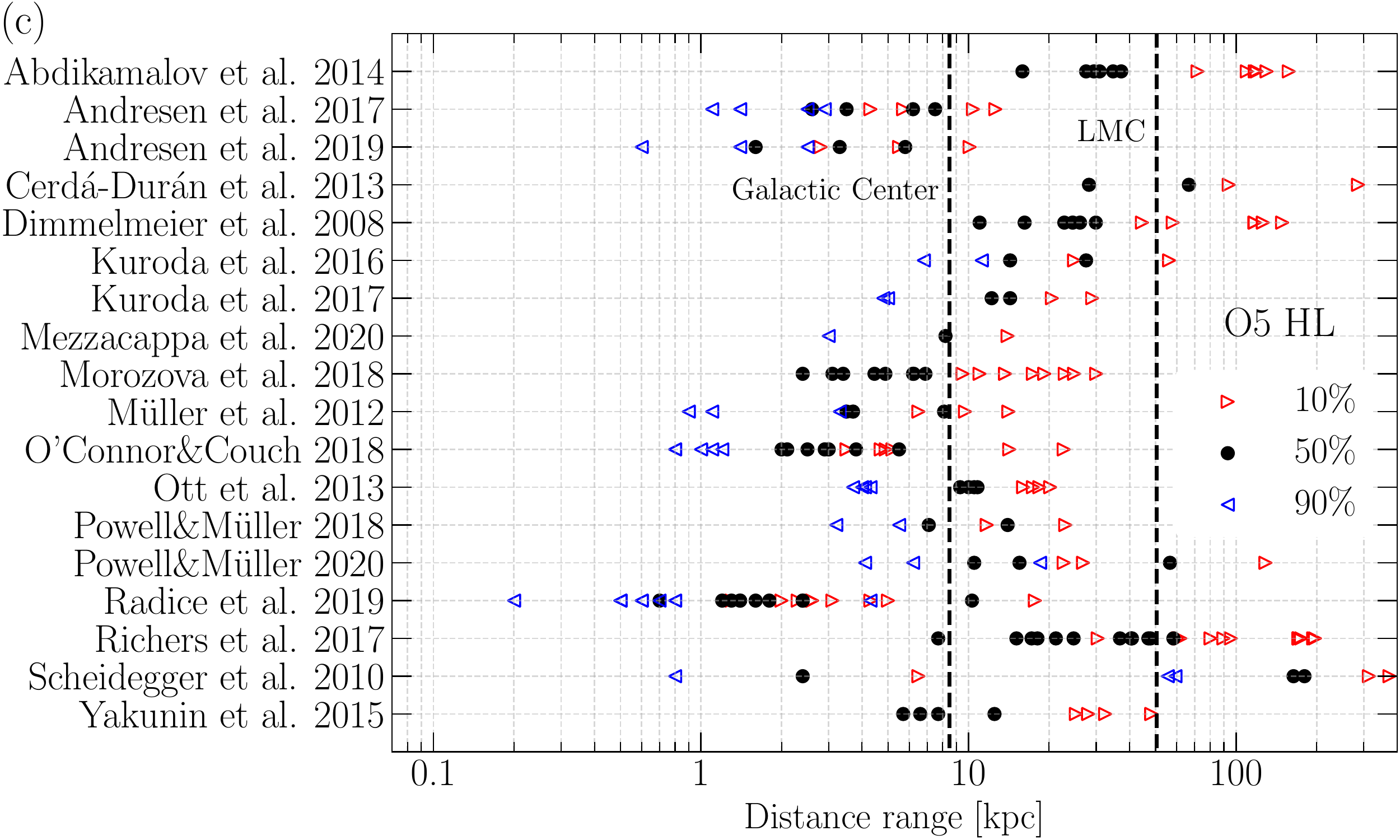}
\caption{Detection efficiency curves for example waveforms are presented in panels (a) and (b). The numbers in the brackets are distances at 50\% detection efficiencies. Panel (c) shows the distances at 10\%, 50\% and 90\% detection efficiencies for all waveforms analyzed in this paper. The predicted detection ranges for O5 are typically between 1\,kpc and 100\,kpc. This range contains the distances to the Galactic center and the Large Magellanic Cloud (LMC) that hosted SN~1987A. The detectability of GW signals coming from 3D simulations can reach almost 100\% detection efficiency at close distances, while linearly polarized waveforms reach only 70\% of the detection efficiency. }
\label{fig:eff_dist}
\end{figure*}


\section{Results}
\label{sec:results}

\begin{figure*}[hbt]
\begin{minipage}{.5\linewidth}
    \centering
    \includegraphics[width=0.96\linewidth]{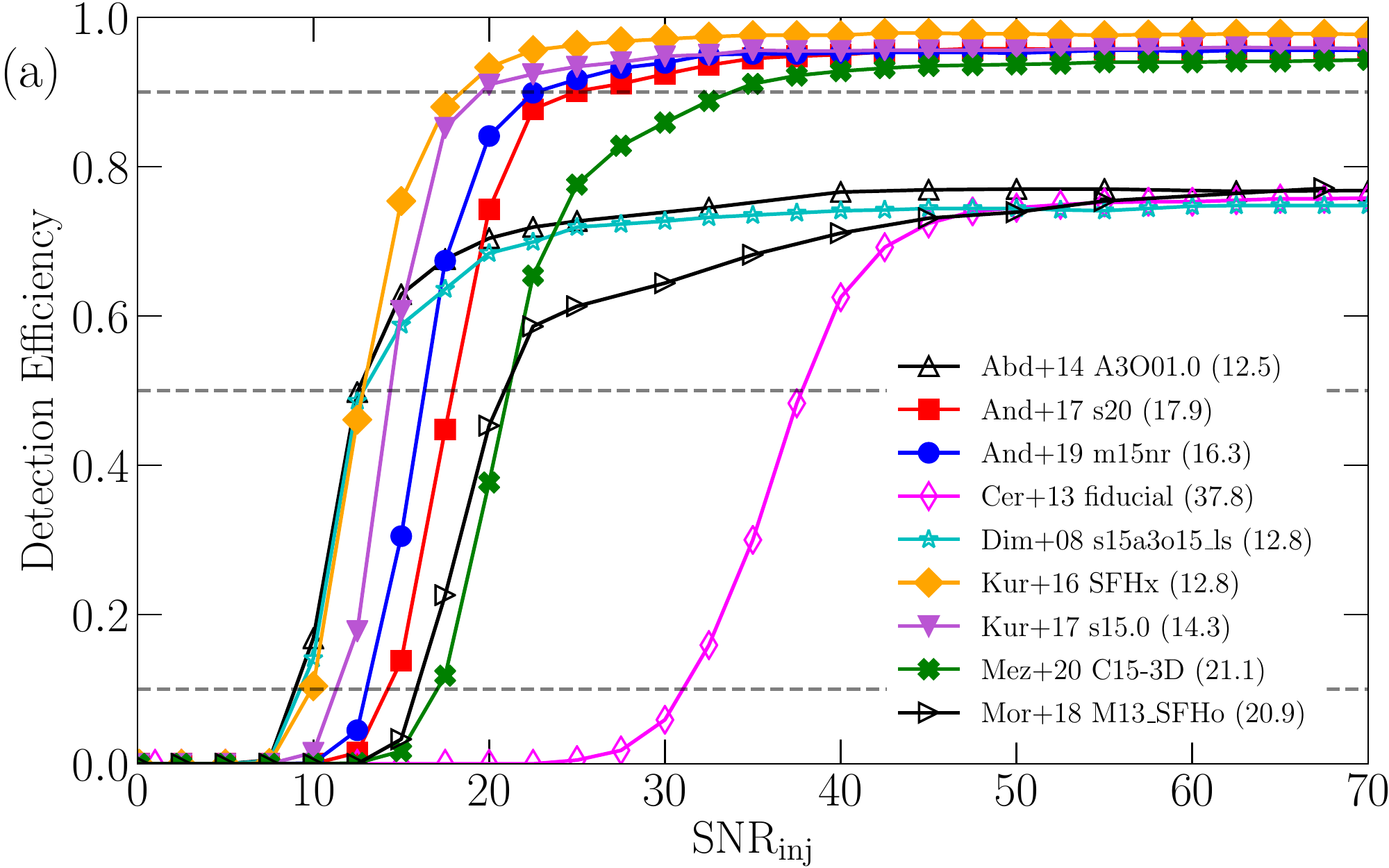}
\end{minipage}%
\begin{minipage}{.5\linewidth}
    \centering
    \includegraphics[width=0.96\linewidth]{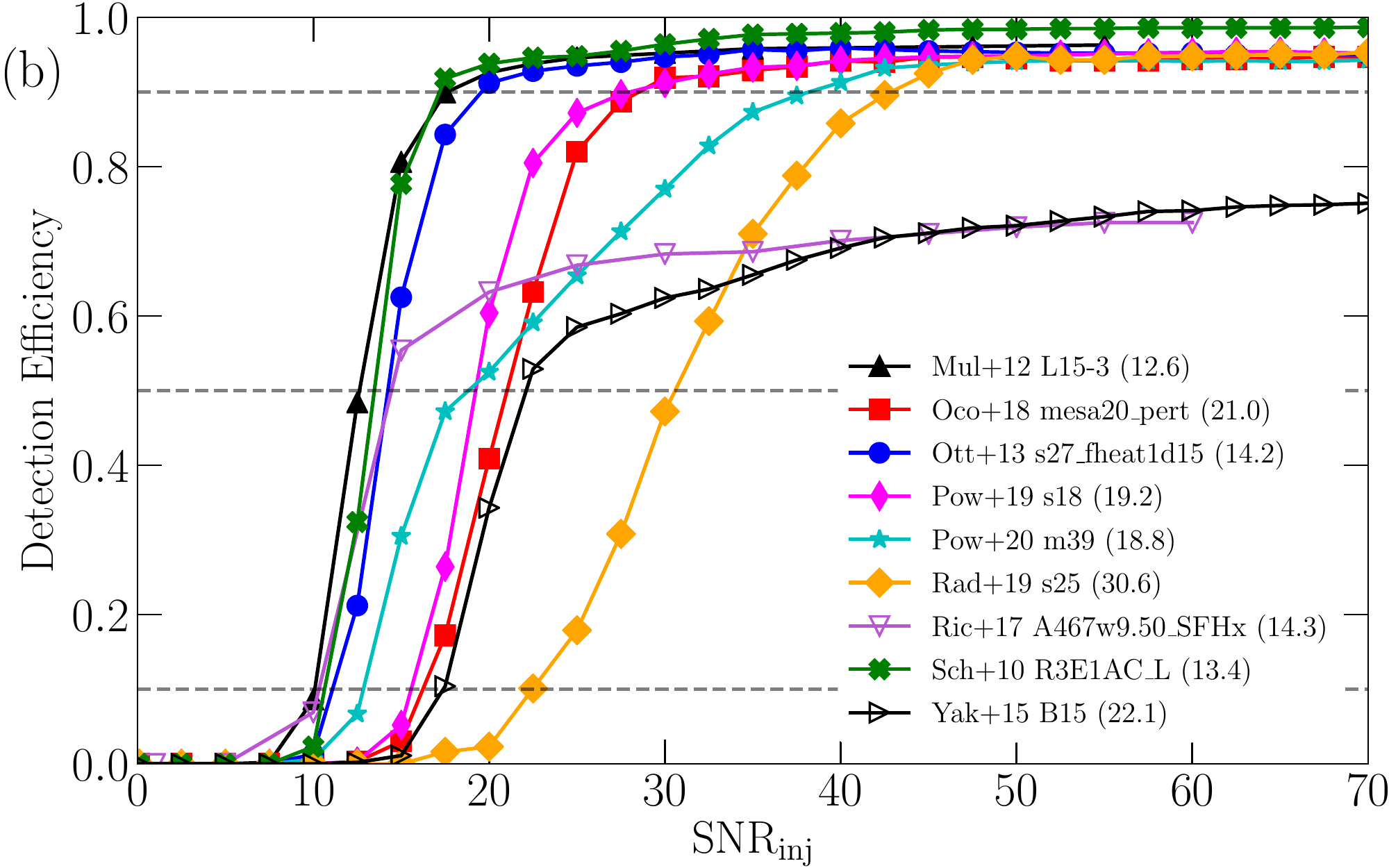}
\end{minipage}\par\medskip
\centering
\includegraphics[width=0.6\linewidth]{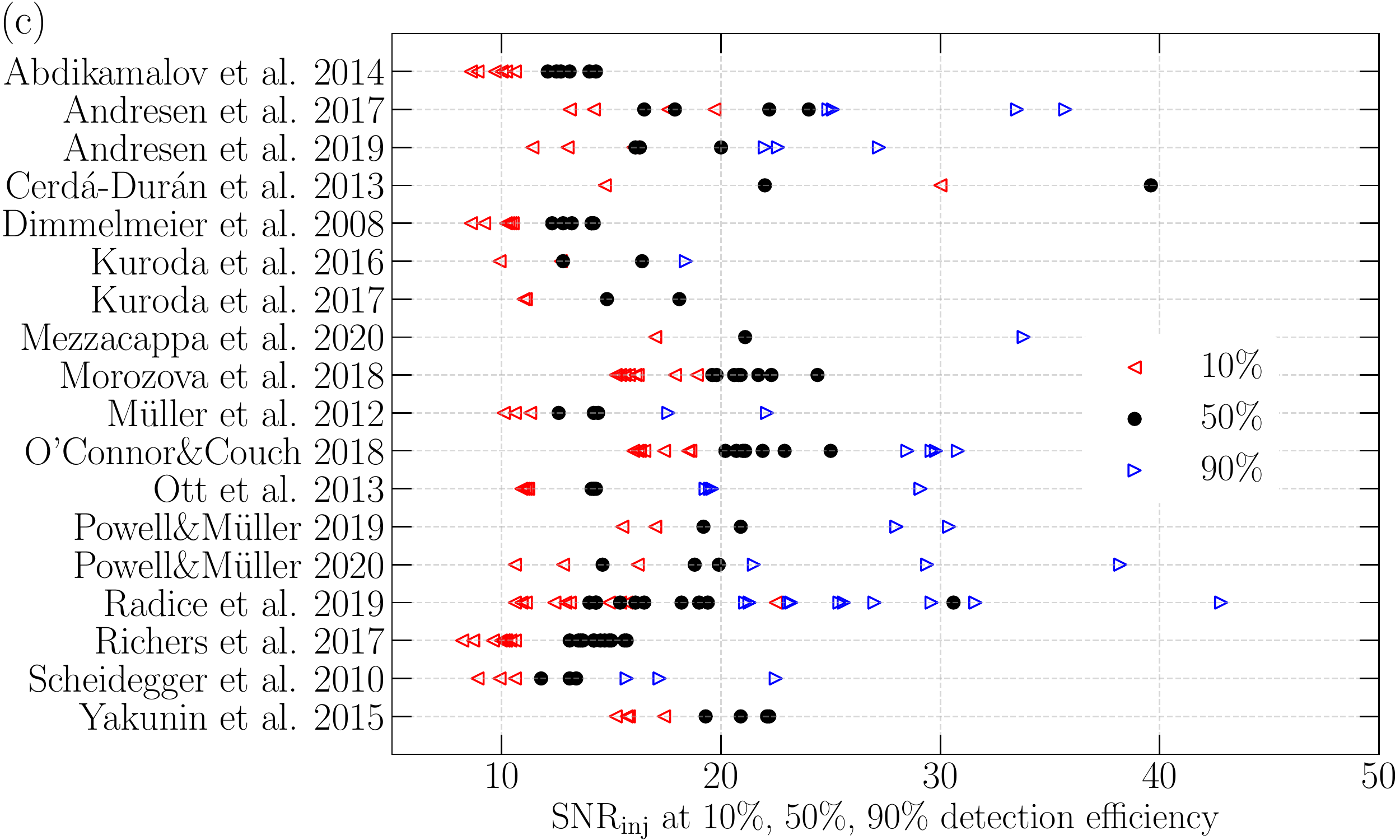}
\caption{Detection efficiency curves as a function of injected SNR ($SNR_\mathrm{inj}$) for example waveforms are presented in panels (a) and (b). The numbers in the brackets are SNRs at 50\% detection efficiencies. Panel (c) shows the SNRs at 10\%, 50\% and 90\% detection efficiencies for all waveforms analyzed in this paper. The waveforms are typically detectable at SNR of 10-25. The Cer+13 fiducial waveform is almost 2\,s long which makes it challenging to detect. The dominant GW emission periods for Rad+19 s25 are 0.6\,s apart, making it challenging to group them by the algorithm. The linearly polarized waveforms reach only 70\% detection efficiency.
}
\label{fig:eff_snr}
\end{figure*}

\subsection{Detection ranges}
\label{sec:det}

The detection ranges for the projected sensitivities of the LIGO detectors in O4 and O5 are presented in Figure~\ref{fig:eff_dist} and Table~\ref{tab:results}. Top panels of Figure~\ref{fig:eff_dist} provide example detection efficiency curves for projected O5 LIGO sensitivities. Bottom panel of Figure~\ref{fig:eff_dist} summarize distances at 10\%, 50\% and 90\% detection efficiencies for all analyzed waveforms. 

In Figure~\ref{fig:eff_dist}, a detection efficiency curve for a certain waveform can be interpreted as the probability of detecting that waveform as a function of the source distance. The numbers in the brackets are the detection ranges (distances at 50\% detection efficiency). The values vary significantly, from around 1\,kpc to over 100\,kpc. The maximum values of the detection efficiency curves for waveforms calculated in 3D simulations are above 90\% while it is around 70\% for linearly polarized GW signals. The HL network used for this analysis is sensitive effectively to only one polarization (the arms of H1 and L1 detectors are approximately parallel). Depending on the polarization angle a waveform may not be registered at the output of the detectors, even if the amplitude is large compared to the noise level. Notably, the best-studied bounce signal has only one polarization component and with 30\% probability, the signal will not be detectable even for a very nearby CCSN.

The bottom panel of Figure~\ref{fig:eff_dist} provides a broad overview of how well the GW signals from CCSNe can be detected in O5. Typically, the detection ranges for the waveforms generated in neutrino-driven explosions are up to around 10\,kpc and only a few GW signals can be detected up to the edge of the Milky Way. When a star explodes according to the MHD-driven mechanism, the detection ranges may exceed the distance of the Large Magellanic Cloud (49.6\,kpc~\cite{2019Natur.567..200P}). The largest detection ranges are obtained for Sch+10 (around 100\,kpc for R3E1AC$_\mathrm{L}$ and R4E1CA$_\mathrm{L}$) and Pow+20 (60\,kpc for y20). These results are in a qualitative agreement with previous studies and conclusions from the optically targeted search performed with O1-O2 data~\cite{Abbott:2019pxc,Gossan:2015xda} where the detection ranges for MHD-driven explosions are much larger than for neutrino-driven explosions. It is worth mentioning that the detection ranges for the MHD-driven explosions could increase significantly if the amplitudes of the turbulent phase (not available for Abd+14, Dim+08, and Ric+17) are comparable with the core bounce one.

Table~\ref{tab:results} summarizes the distances at 10\%, 50\% and 90\% detection efficiencies for waveforms described in Table~\ref{tab:models}. The LIGO detectors will be improved between O4 and O5 consistently in a large frequency range of a factor $\sim1.8$ (see Figure~\ref{fig:recolor}) and the detection ranges improve by around the same factor. 

\begin{figure*}[hbt]
 \begin{minipage}[c][][t]{0.495\textwidth}
    \vspace*{\fill}
    \flushleft
    \includegraphics[width=0.96\linewidth]{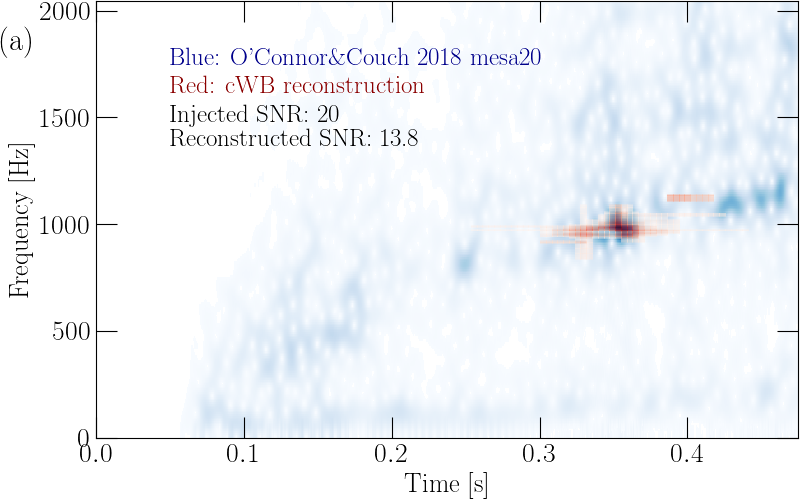}
 \end{minipage}%
 \begin{minipage}[c][][t]{0.495\textwidth}
    \vspace*{\fill}
    \flushright
    \includegraphics[width=0.96\linewidth]{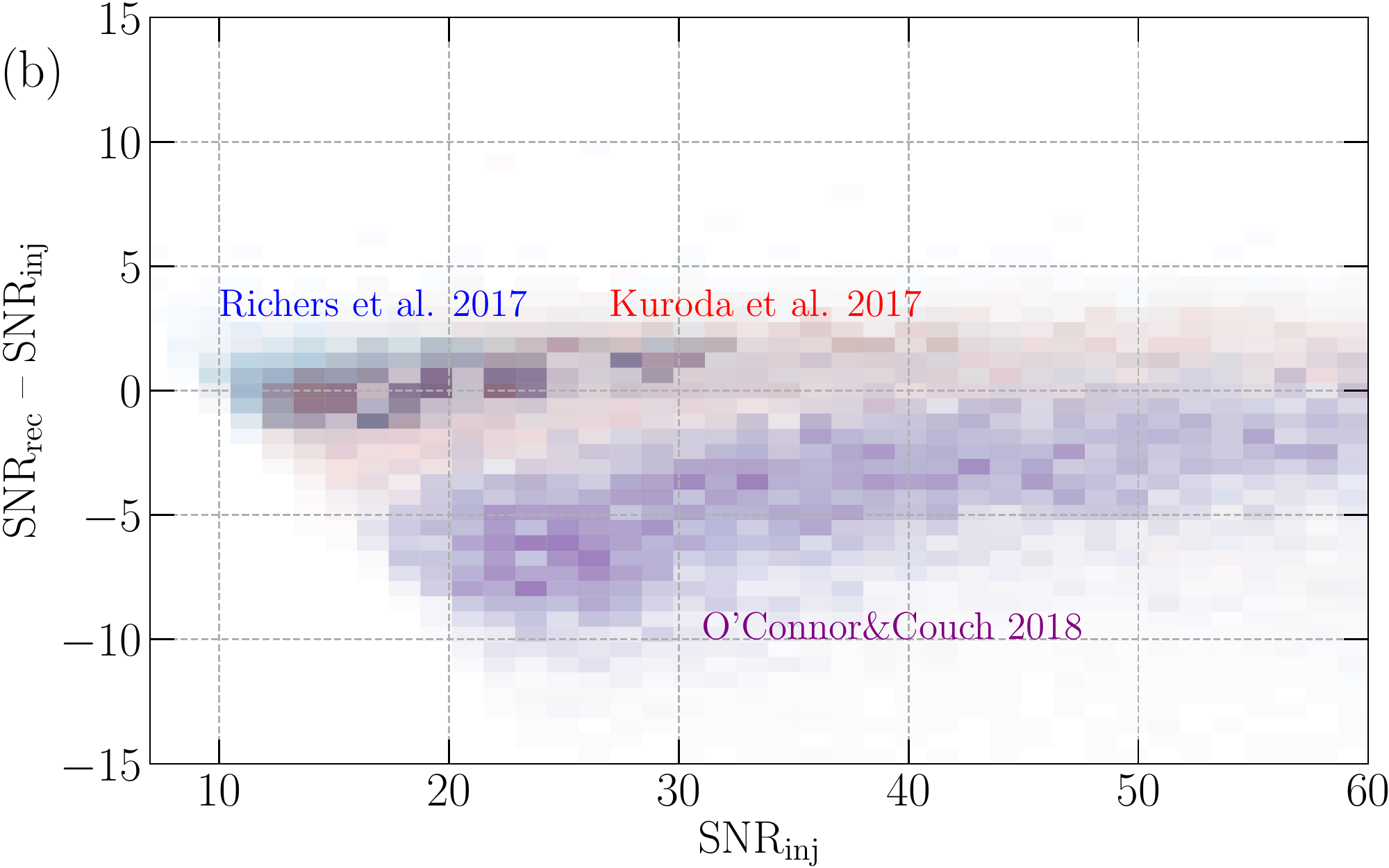}
 \end{minipage}
	\caption{
	Panel (a) shows an example Oco+18 mesa20 waveform with an overlay of the cWB reconstruction. The Waveform is relatively long and broadband and for small SNR values the algorithm does not reconstruct a large part of the signal. Panel (b) quantifies the SNR difference as a function of injected SNR. The cWB search reconstructs well the signals that are short in duration and its capabilities decrease with the increasing signal complexity in their time-frequency evolution.}
	\label{fig:snr}
\end{figure*}

\subsection{Minimum detectable SNR}
\label{sec:snr}

The cWB algorithm is sensitive to a wide range of GW signals but it is not equally sensitive to all morphologies. In general, waveforms that are short and narrowband are easier to detect than waveforms that are long, broadband, or fragmented in the time-frequency domain. As an illustration, binary BH signals usually have a continuous evolution in the time-frequency domain, in the LIGO band they are typically relatively short and narrowband. On the contrary, the waveforms from CCSNe often have very complex signatures in time and frequency. For example, the peak frequencies of GWs from PNS oscillations evolve from around 100\,Hz up to a few kHz during the first second after the collapse. The time-frequency evolution of these oscillations often is not continuous and depends on the amount of accreting matter. Moreover, rapid plumes of infalling matter can cause the generation of a broadband GW signal. Additionally, the GWs from SASI/convection and the PNS oscillations can be disconnected in the time-frequency domain.

\begin{figure}[t] 
\includegraphics[width=0.99\linewidth]{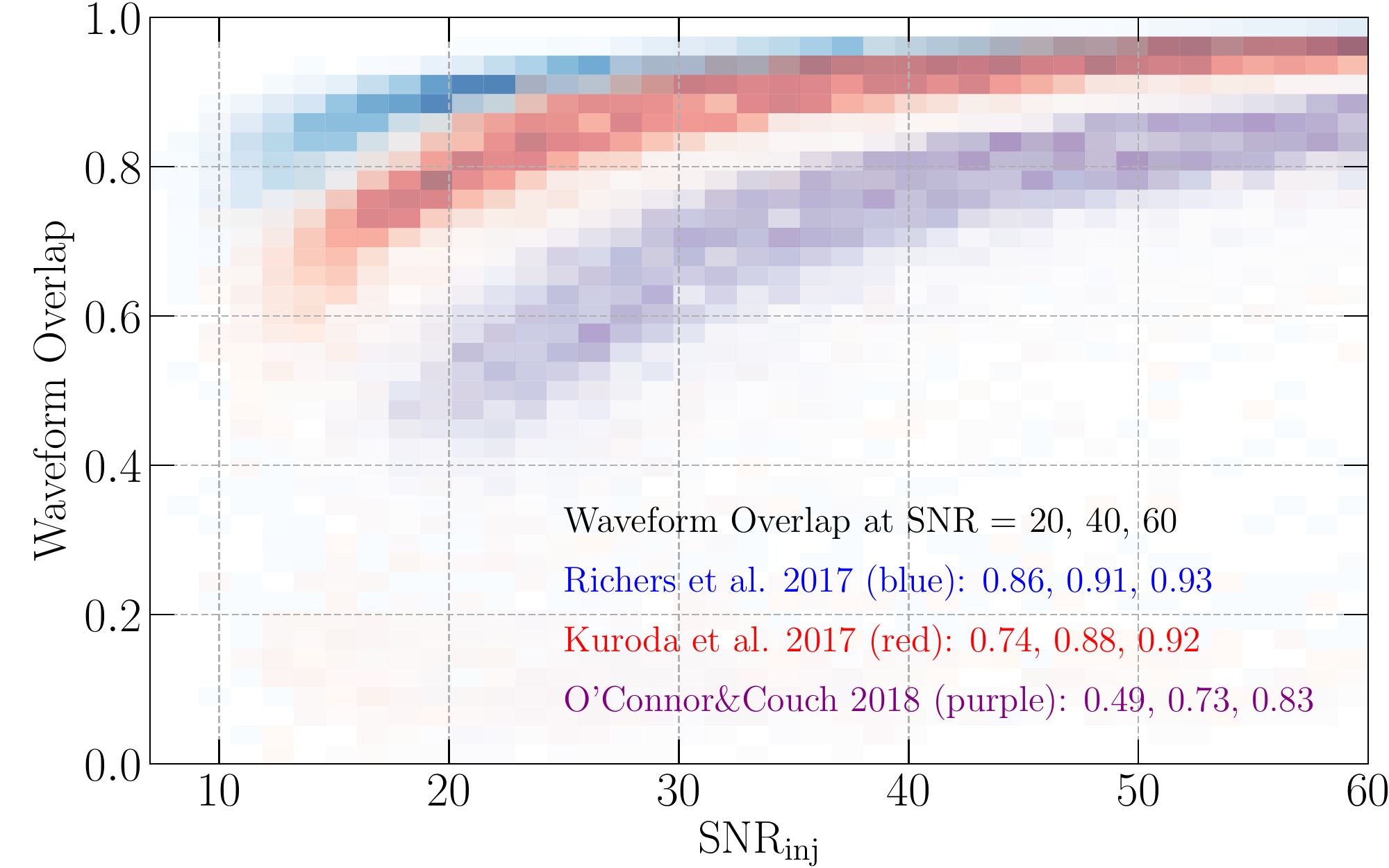}
\caption{Waveform overlap as a function of the injected SNR. The waveforms that are short, like Rad+17 are accurately reconstructed for small SNRs, while waveforms that are long and broadband, like Oco+18, need to be relatively strong to be reconstructed accurately.}
\label{fig:ff}
\end{figure} 

The top panels of Figure~\ref{fig:eff_snr} present detection efficiency curves as a function of injected SNR for projected O5 sensitivity and an HL network. The numbers in the brackets are the minimum detectable SNR (SNR at 50\% detection efficiency). The bottom panel of Figure~\ref{fig:eff_snr} summarizes the SNR values at 10\%, 50\%, and 90\% detection efficiency for all analyzed waveforms. The minimum detectable SNR is typically in the range of 10-25. The smallest values are reported for short waveforms ($<$200\,ms) such as Abd+14, Dim+08, Kur+17, Ott+13, Sch+10, and Ric+17, or when they are narrowband, e.g. Mul+12. The minimum detectable SNR is increasing with the complexity of the waveform morphology and the GW signals from neutrino-driven explosions have higher minimum detectable SNR values, such as for And+16, Mor+18, Oco+18, and Rad+19. The highest minimum detectable SNR is given for the Cer+13 fiducial waveform. This signal represents a BH formation after almost 2\,s with a broadband spectrum making it challenging to detect. If the star collapses to a BH faster (e.g. Cer+13~slow or~\cite{Chan2018,Kuroda2018,Pan2018,Pan2020,2021arXiv210106889P}) then the corresponding SNR to capture the full signal is smaller.
Similar to the results obtained in Section~\ref{sec:det}, the detection efficiency for linearly polarized waveforms do not exceed around 70\% detection efficiency.

Table~\ref{tab:results} provides the SNR values at 10\%, 50\% and 90\% detection efficiency for the waveforms listed in Table~\ref{tab:models}. Given that the predicted improvement between O4 and O5 is uniform across the LIGO frequency band, the results obtained in both observing runs are practically the same.

The challenge of detecting and reconstructing GW signals that are long and broadband is illustrated in the left panel of Figure~\ref{fig:snr}. The plot shows a spectrogram of the Oco+18 mesa20 waveform and an overlay of a cWB reconstruction. The signal is almost 0.5\,s long and its energy spans up to 2\,kHz. The waveform has a visible SASI/convection signature and the peak frequency evolution of the PNS oscillations. The overall energy of the signal is spread rather uniformly in the time-frequency domain. In this particular case, the injected SNR is 20 while the cWB reconstructed SNR is 13.8. The peak frequency is around 1\,kHz, and only part of the signal around the peak frequency is reconstructed. This situation is typical for the waveforms from neutrino-driven explosions.


\begin{table*}

\caption{
The results presenting the sensitivity of cWB to the detection of GWs from a variety of CCSN models. The predicted detection ranges for O4 and O5 are calculated at 10\%, 50\% and 50\% detection efficiency. The detectable SNR is also calculated at 10\%, 50\% and 90\% detection efficiency. The waveform overlap (accuracy of cWB reconstruction) is an averaged at injected SNR of 20, 40 and 60.
  }
\begin{tabular*}{\textwidth} {@{\extracolsep{\textwidth minus \textwidth}} cccccccccccccc}
\hline
\hline
\multicolumn{1}{c}{Waveform}
&\multicolumn{1}{c}{Waveform}
&\multicolumn{3}{c}{O4 det. range [kpc]}
&\multicolumn{3}{c}{\phantom{0}O5 det. range [kpc]}
&\multicolumn{3}{c}{Detect. SNR}
&\multicolumn{3}{c}{Wav. Overlap}
\\
Family 
& Identifier
& 90\% & 50\% & 10\%
& 90\% & 50\% & 10\%
& 10\% & 50\% & 90\%
& 20   & 40   & 60 \\
\hline
\hline
\multirow{4}{*}{\makecell{Abdikamalov\\ et al. 2014 \cite{Abdikamalov_2014}}}
&A1O01.0 & NaN & 15.9 & 58.7    & NaN & 29.4 & 109.7    & 9.7 & 12.7 & NaN     & 0.83 & 0.90 & 0.93  \\
&A2O01.0 & NaN & 19.3 & 71.0    & NaN & 35.2 & 130.0    & 10.0 & 13.1 & NaN    & 0.88 & 0.93 & 0.94  \\
&A3O01.0 & NaN & 20.1 & 84.6    & NaN & 37.1 & 157.4    & 8.9 & 12.5 & NaN     & 0.86 & 0.92 & 0.95  \\
&A4O01.0 & NaN & 8.4  & 39.3    & NaN & 15.2 & 72.3     & 10.2 & 14 & NaN      & 0.88 & 0.91 & 0.94  \\
\hline
\multirow{4}{*}{\makecell{Andresen\\ et al. 2017 \cite{Andresen_2017}}}
& s11 & 0.6 & 1.4 & 2.3   & 1.1 & 2.6 & 4.3     & 13.1 & 16.5 & 25.1     & 0.59 & 0.82 & 0.88  \\
& s20 & 1.4 & 3.4 & 5.6   & 2.5 & 6.2 & 10.4    & 14.2 & 17.9 & 24.9     & 0.50 & 0.79 & 0.88  \\
& s20s & 1.6 & 4.1 & 6.8   & 2.9 & 7.5 & 12.6    & 19.7 & 24.0 & 35.7     & 0.35 & 0.71 & 0.84  \\
& s27 & 0.8 & 1.9 & 3.1   & 1.4 & 3.5 & 5.7     & 17.6 & 22.2 & 33.5     & 0.71 & 0.68 & 0.83  \\
\hline
\multirow{3}{*}{\makecell{Andresen\\ et al. 2019 \cite{Andresen_2019}}}
& m15fr & 1.4 & 3.2 & 5.6    & 2.5 & 5.8 & 10.1    & 11.4 & 16.1 & 22.0     & 0.61 & 0.77 & 0.85  \\
& m15nr & 0.8 & 1.8 & 3.1    & 1.4 & 3.3 & 5.5     & 13.0 & 16.3 & 22.6     & 0.59 & 0.82 & 0.88  \\
& m15r  & 0.3 & 0.9 & 1.5    & 0.6 & 1.6 & 2.8     & 16.0 & 20.0 & 27.6     & 0.46 & 0.78 & 0.86  \\
\hline
\multirow{2}{*}{\makecell{Cerd\'a-Dur\'an\\ et al. 2013 \cite{Cerda-Duran2013}}}
& fiducial & NaN & 15.7 & 51.5     & NaN & 28.2 & 93.9      & 31.0 & 37.8 & NaN     & 0.52 & 0.81 & 0.87  \\
& slow     & NaN & 35.9 & 154.3    & NaN & 66.6 & 285.7     & 15.3 & 19.7 & NaN     & 0.35 & 0.63 & 0.81  \\
\hline
\multirow{3}{*}{\makecell{Dimmelmeier\\ et al. 2008 \cite{Dimmelmeier2008}}}
& s15A2O09-ls & NaN & 14.5 & 60.1     & NaN & 26.1 & 117.5     & 10.2 & 13.2 & NaN      & 0.86 & 0.91 & 0.93  \\
& s15A3O15-ls & NaN & 13.6 & 59.4     & NaN & 24.5 & 117.1     &  9.2 & 12.8 & NaN      & 0.90 & 0.94 & 0.95  \\
& s20A3O09-ls & NaN & 12.5 & 59.9     & NaN & 22.8 & 125.9      & 10.3 & 14.2 & NaN      & 0.84 & 0.90 & 0.92  \\
\hline
\multirow{2}{*}{\makecell{Kuroda\\ et al. 2016 \cite{Kuroda2016}}}
& SFHx & 4.9 & 11.8 & 23.8    & 8.7 & 21.6 & 43.3     &  10.4 & 14.1 & 22.1     & 0.63 & 0.82 & 0.88  \\
& TM1  & 3.7 &  8.0 & 13.2    & 6.5  & 14.5 & 24.8     & 12.7 & 15.5 & 19.5     & 0.61 & 0.82 & 0.88  \\
\hline
\multirow{2}{*}{\makecell{Kuroda\\ et al. 2017 \cite{Kuroda2017}}}
& s11.2 & 2.5 & 7.7 & 15.9    & 4.8 & 14.3 & 29.0   & 10.0 & 12.7 & 21.3     & 0.82 & 0.90 & 0.93  \\
& s15.0 & 2.7 & 6.7 & 11.7    & 5.0 & 12.2 & 20.5   & 11.2 & 14.3 & 19.5     & 0.75 & 0.89 & 0.92  \\
\hline
\multirow{2}{*}{\makecell{Mezzacappa\\ et al. 2020 \cite{Mezzacappa:2020lsn}}} &
\multirow{2}{*}{\makecell{c15-3D}} & 
\multirow{2}{*}{\makecell{1.8}} & 
\multirow{2}{*}{\makecell{4.4}} & 
\multirow{2}{*}{\makecell{7.4}} & 
\multirow{2}{*}{\makecell{3.0}} & 
\multirow{2}{*}{\makecell{8.2}} & 
\multirow{2}{*}{\makecell{14.0}} & 
\multirow{2}{*}{\makecell{17.0}} & 
\multirow{2}{*}{\makecell{21.1}} & 
\multirow{2}{*}{\makecell{33.8}} & 
\multirow{2}{*}{\makecell{0.42}} & 
\multirow{2}{*}{\makecell{0.69}} & 
\multirow{2}{*}{\makecell{0.82}} \\
&  &  &  &     &  &  &      &  &  &     &  & &  \\
\hline
\multirow{4}{*}{\makecell{Morozova\\ et al. 2018 \cite{Morozova2018}}}
& M10\_LS220 & NaN & 1.3 & 5.2    & NaN & 2.4 & 9.5     & 16.2 & 21.7 & NaN     & 0.47 & 0.72 & 0.81  \\
& M10\_DD2   & NaN & 1.9 & 7.4    & NaN & 3.4 & 13.7     & 15.2 & 19.6 & NaN     & 0.57 & 0.80 & 0.85  \\
& M13\_SFHo  & NaN & 2.3 & 10.2    & NaN & 4.5 & 19.2     & 15.8 & 20.9 & NaN     & 0.49 & 0.74 & 0.80  \\
& M19\_SFHo  & NaN & 3.9 & 16.7    & NaN & 6.9 & 30.0     & 18.9 & 24.4 & NaN     & 0.37 & 0.68 & 0.78  \\
\hline
\multirow{3}{*}{\makecell{M\"uller\\ et al. 2012 \cite{Muller:2012}}}
& L15-3 & 1.7 & 4.3 & 8.0    & 3.3 & 8.1 & 14.1     & 10.1 & 12.6 & 17.6     & 0.73 & 0.81 & 0.84  \\
& N20-2 & 0.5 & 1.9 & 3.6    & 1.1 & 3.5 &  6.5     & 11.3 & 14.4 & 22.1     & 0.68 & 0.79 & 0.84  \\
& W15-4 & 0.5 & 1.9 & 5.2    & 0.9 & 3.7 &  9.7     & 10.6 & 14.2 & 42.2     & 0.71 & 0.83 & 0.88  \\
\hline
\multirow{4}{*}{\makecell{O'Connor\&Couch\\ 2018 \cite{OConnor2018}}}
& mesa20        & 0.4 & 1.1 & 1.9    & 0.8 & 2.0 & 3.5     & 16.3 & 20.7 & 30.8     & 0.50 & 0.70 & 0.82  \\
& mesa20\_LR    & 0.6 & 1.4 & 2.6    & 1.0 & 2.5 & 4.7     & 18.5 & 25.0 & 42.3     & 0.45 & 0.67 & 0.79  \\
& mesa20\_pert  & 0.7 & 1.6 & 2.9    & 1.2 & 2.9 & 4.9     & 16.2 & 21.0 & 28.5     & 0.47 & 0.75 & 0.84  \\
& mesa20\_v\_LR & 0.4 & 1.1 & 1.9    & 0.8 & 2.1 & 3.5     & 16.0 & 20.2 & 29.6     & 0.51 & 0.78 & 0.87  \\
\hline
\multirow{4}{*}{\makecell{Ott\\ et al. 2013 \cite{Ott2013}}}
& s27-fheat1.00 & 2.4 & 5.8 & 10.5    & 4.3 & 10.8 & 20.2      & 11.1 & 14.3 & 20.1     & 0.75 & 0.89 & 0.92  \\
& s27-fheat1.05 & 2.0 & 5.8 & 10.6    & 4.1 & 10.5 & 18.4      & 10.9 & 14.1 & 19.3     & 0.74 & 0.88 & 0.91  \\
& s27-fheat1.10 & 2.4 & 5.8 & 10.0    & 4.0 & 10.0 & 17.4      & 11.2 & 14.2 & 19.6     & 0.75 & 0.88 & 0.92  \\
& s27-fheat1.15 & 1.9 & 5.2 &  9.0    & 3.7 &  9.3 & 16.0      & 11.0 & 14.2 & 19.5     & 0.76 & 0.88 & 0.92  \\
\hline
\multirow{2}{*}{\makecell{Powell\&M\"uller\\ 2019 \cite{Powell2019}}}
& s3.5\_pns & 1.8 & 3.9 &  6.4    & 3.2  &  7.1 & 11.7     & 17.0 & 20.9 & 30.4     & 0.44 & 0.75 &  0.83 \\
& s18   & 3.2 & 7.7 & 12.7    & 5.5 & 14.0 & 23.0     & 15.5 & 19.2 & 28.0     & 0.47 & 0.73 & 0.81  \\
\hline
\multirow{3}{*}{\makecell{Powell\&M\"uller\\ 2020 \cite{Powell2020}}}
& m39 & 10.3 & 30.7 & 70.2     & 18.5 & 56.6 & 128.8    & 12.8 & 18.8 & 38.2     & 0.57 & 0.73 & 0.81  \\
& s18np & 2.3 & 5.7  & 12.3     &  4.1 & 10.5 &  22.7    & 10.6 & 14.6 & 21.5     & 0.67 & 0.81 & 0.88  \\
& y20 & 3.4 & 8.5  & 14.6     &  6.2 & 15.5 &  26.8    & 16.2 & 19.9 & 29.4     & 0.42 & 0.72 & 0.82  \\
\hline
\multirow{3}{*}{\makecell{Radice\\ et al. 2019 \cite{Radice2019}}}
& s9  & 0.0 & 0.4 & 0.7    & 0.2 & 0.7 & 1.3      & 11.1 & 14.3 & 23.1     & 0.73 & 0.84 & 0.91  \\
& s13 & 0.4 & 1.0 & 1.8    & 0.7 & 1.8 & 3.1      & 10.9 & 14.3 & 21.1     & 0.68 & 0.80 & 0.87  \\
& s25 & 2.4 & 5.6 & 9.4    & 4.3 & 10.3 & 17.7     & 22.5 & 30.6 & 42.8      & 0.43 & 0.65 & 0.78  \\
\hline
\multirow{4}{*}{\makecell{Richers\\ et al. 2017 \cite{Richers2017}}}
& A467\_w0.50\_SFHx  & NaN & 8.0 & 32.9    & NaN & 15.1 & 60.6     & 8.70 & 13.7 & NaN     & 0.86 & 0.91 & 0.93 \\
& A467\_w0.50\_LS220 & NaN & 10.3 & 43.0    & NaN & 18.1 & 80.3     & 10.2 & 14.2 & NaN     & 0.85 & 0.93 &  0.94 \\
& A467\_w9.50\_SFHx  & NaN & 24.2 & 105.2     & NaN & 47.9 & 194.2      & 10.3 & 14.3 & NaN     & 0.82 & 0.91 & 0.91  \\
& A467\_w9.50\_LS220 & NaN & 22.5 & 90.5      & NaN & 40.8 & 171.9      & 10.1 & 14.1 & NaN     & 0.82 & 0.89 &  0.92 \\
\hline
\multirow{3}{*}{\makecell{Scheidegger\\ et al. 2010 \cite{Scheidegger2010}}}
& R1E1CA\_L & 0.4 & 1.3 & 3.5        & 0.8 & 2.4 & 6.5          & 9.9  & 13.1 & 22.5     & 0.76 & 0.86 & 0.91 \\
& R3E1AC\_L & 29.9 & 89.6 & 171.8    & 55.5 & 163.9 & 313.9     & 10.6 & 13.4 & 17.2     & 0.76 & 0.89 & 0.93 \\
& R4E1FC\_L & 31.8 & 98.4 & 203.4    & 59.3 & 180.1 & 374.6     & 8.9  & 11.8 & 15.7     & 0.81 & 0.91 & 0.94 \\
\hline
\multirow{4}{*}{\makecell{Yakunin\\ et al. 2015 \cite{Yakunin2015a}}}
& B12 & NaN & 3.6 & 13.6    & NaN &  6.6 & 25.2     & 15.2 & 19.3 & NaN     & 0.51 & 0.80 & 0.88  \\
& B15 & NaN & 4.3 & 17.9    & NaN &  7.7 & 32.4     & 17.4 & 22.1 & NaN     & 0.44 & 0.78 & 0.87  \\
& B20 & NaN & 3.0 & 15.2    & NaN &  5.7 & 28.2     & 15.8 & 22.2 & NaN     & 0.52 & 0.82 & 0.89  \\
& B25 & NaN & 6.6 & 26.1    & NaN & 12.5 & 48.2     & 15.7 & 20.9 & NaN     & 0.49 & 0.76 & 0.85  \\
\hline
\hline
\end{tabular*}
\label{tab:results}
\end{table*}


\begin{figure*}[hbt] 
\includegraphics[width=0.70\linewidth]{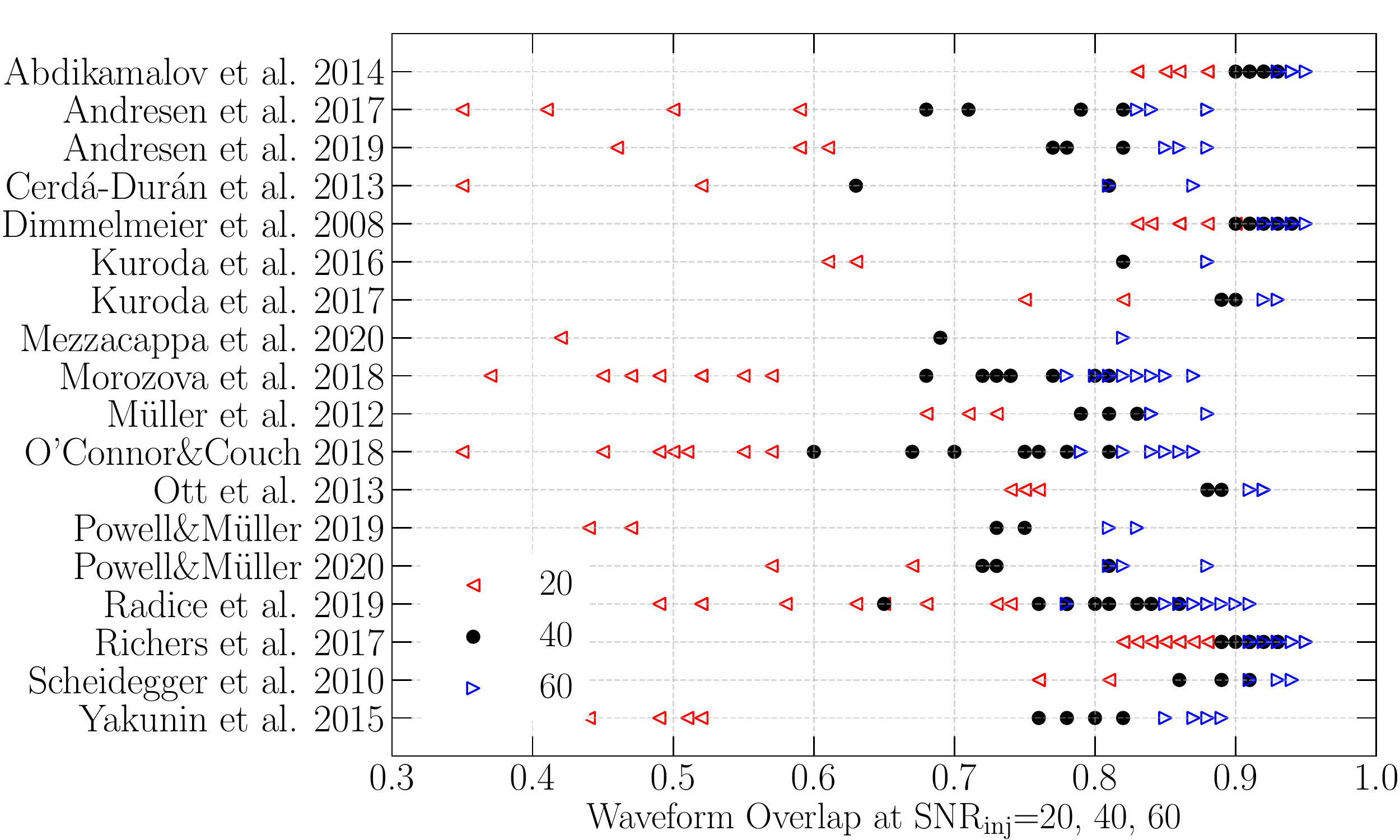}
\caption{Waveform overlaps (reconstruction accuracy) for all waveforms analyzed in this paper at injected SNR of 20, 40 and 60. The accuracy of reconstruction is close to unity for short waveforms such as Abd+14, Dim+08, Ric+17. It decreases with waveform length and its complexity in the time-frequency evolution. }
\label{fig:set-ff}
\end{figure*}

The right panel of Figure~\ref{fig:snr} presents the distributions of the difference between the reconstructed and injected SNR as a function of the injected SNR for three waveform families. Ric+17 simulate the GW bounce signals and the waveforms are only 6\,ms long. The minimum detectable SNR is around 10 and the reconstructed SNR matches well the injected SNR. The Kur+17 waveforms are shorter than 200\,ms, they lack emission from the PNS oscillations and the peak frequency is below 400\,Hz. In this case, the reconstructed SNR is close to the injected SNR, but the difference is larger than for the Rich+17 waveforms. The Oco+18 waveforms have an even more complex time-frequency structure and the reconstructed SNR is significantly underestimated. It is worth mentioning that it is possible to improve the detectability of long and broadband waveforms by decreasing the cWB internal thresholds. However, it comes with the expense of increasing significantly the number of background events and diminishing the significance of the detected GW signals.

\subsection{Reconstruction accuracy}

To quantify the accuracy of the cWB reconstruction, we use the waveform overlap, or a  match~\cite{Abbott:2020tfl}, between a detected $\mathbf{w} = \{w_{k}(t)\}$ and injected waveform $\mathbf{h} = \{{h}_{k}(t)\}$:
\begin{equation}
 \label{eqn:ff}
 \mathrm{O}(\mathbf{w},\mathbf{h}) =  \frac{(\mathbf{w}|\mathbf{h}) }{\sqrt{(\mathbf{w}|\mathbf{w}) } \sqrt{(\mathbf{h}|\mathbf{h}) } } \,.
\end{equation}
The scalar product $(.|.)$ is defined in the time domain as  
\begin{equation}
\label{sprod}
(\mathbf{w}|{\mathbf{h}}) = \sum_k \int_{t_{1}}^{t_{2}} {w}_k(t) {h}_k(t) dt,
\end{equation}
where the index $k$ is a detector number and $[t_1, t_2]$ is the time range of the reconstructed event. The waveform overlap ranges from -1 (sign mismatch) to 1 (perfect reconstruction).

The complexity of the GW signals has an impact on the detectable SNR and the corresponding reconstruction accuracy. It is illustrated in Figure~\ref{fig:ff} that shows distributions of the waveform overlaps as a function of the injected SNR for three waveform families. The difference between reconstructed and injected SNR for Ric+17 waveforms is small and they are reconstructed the most accurately, waveform overlaps are around 0.9 even for low SNR signals. The Kur+17 waveforms are longer and the mismatch is larger than for Ric+17 waveforms. The SNR for Oco+17 waveforms is significantly underestimated resulting in relatively low waveform overlap values. The Oco+17 waveforms require a minimum detectable SNR of around 20 and the corresponding average waveform overlap is 0.49. Even for stronger signals, the waveform overlaps do not reach 0.9 on average.

Table~\ref{tab:results} summarizes the waveform overlap values at injected SNRs of 20, 40, and 60 for the waveforms listed in Table~\ref{tab:models}, and Figure~\ref{fig:set-ff} shows results for all analyzed waveforms. The best reconstruction accuracy is obtained for the signals that are very short (Abd+14, Dim+08, and Ric+17). The reconstruction accuracy decreases with waveform length and their complexities, and it is lowest for the waveforms from the neutrino-driven explosions.

\subsection{Reconstruction of GW features}

In the previous section, we quantified how well cWB reconstructs the whole waveforms. Here, we provide a brief qualitative description of how well particular GW features can be reconstructed. When the GW signals are weak cWB usually detects the signal's components around their peak frequency. However, the frequency-dependent detector noise may alter the $f_\mathrm{peak}$ of the injected and reconstructed waveforms. For example, the And+19 m15fr waveform peaks at 711\,Hz and contains a very strong low-frequency SASI component in the frequency range where the GW detectors are most sensitive. As a result, at SNR of 20, the peak frequency reconstructed by cWB is usually around 100\,Hz. This discrepancy can be explain using Figure~\ref{fig:hchar}. The amplitude ratio between the characteristic strain and O4-O5 detector sensitivities are comparable at low and high frequencies, and the reconstructed peak frequency can be ambiguous. Such discrepancies are rare but they happen when the signal is weak.

The peak frequencies of the GW signals usually correspond to the dominant emission processes, as is illustrated in the reconstruction of the Oco+18 waveform in the left panel of Figure~\ref{fig:snr}. When the GW signals are weak, the peak frequency of the dominant GW emission is reconstructed and when the SNR increases, the other parts of the signal get reconstructed as well. For most of the waveforms from the neutrino-driven explosions, the PNS oscillations dominate the waveform amplitudes and the minimum detectable SNR is around 20. The time-frequency path of the increasing peak frequency of the PNS oscillations becomes visible around SNR 30-40. At this SNR level, the SASI/convection becomes reconstructed as well.

The GWs dominated by SASI/convection are relatively narrowband and the minimum detectable SNR is usually smaller for waveforms dominated by PNS oscillations. Examples are Mul+12 or Pow+20 s18 waveforms, their minimum detectable SNR is around 15. Similarly, the prompt-convection signatures are short and they are reconstructed accurately even for small SNR values. For example, Ott+13 and Kur+17 s11 waveforms are dominated by a strong prompt-convection and the minimum detectable SNR is around 15. The GWs from rapidly rotating progenitor stars (Abd+14, Dim+08, Ric+17) are dominated by the bounce and prompt-convection, these waveforms are very short and they are detectable in the SNR around 15 or less. The reconstruction of the BH formation signals that may be long is the most challenging. The duration of Cer+13 fiducial waveform is almost 2\,s and for such long signals, an SNR of 50 is needed to capture the full evolution including the final BH formation signal.

\begin{figure*}[hbt]
 \begin{minipage}[c][][t]{0.495\textwidth}
    \vspace*{\fill}
    \flushleft
    \includegraphics[width=0.96\linewidth]{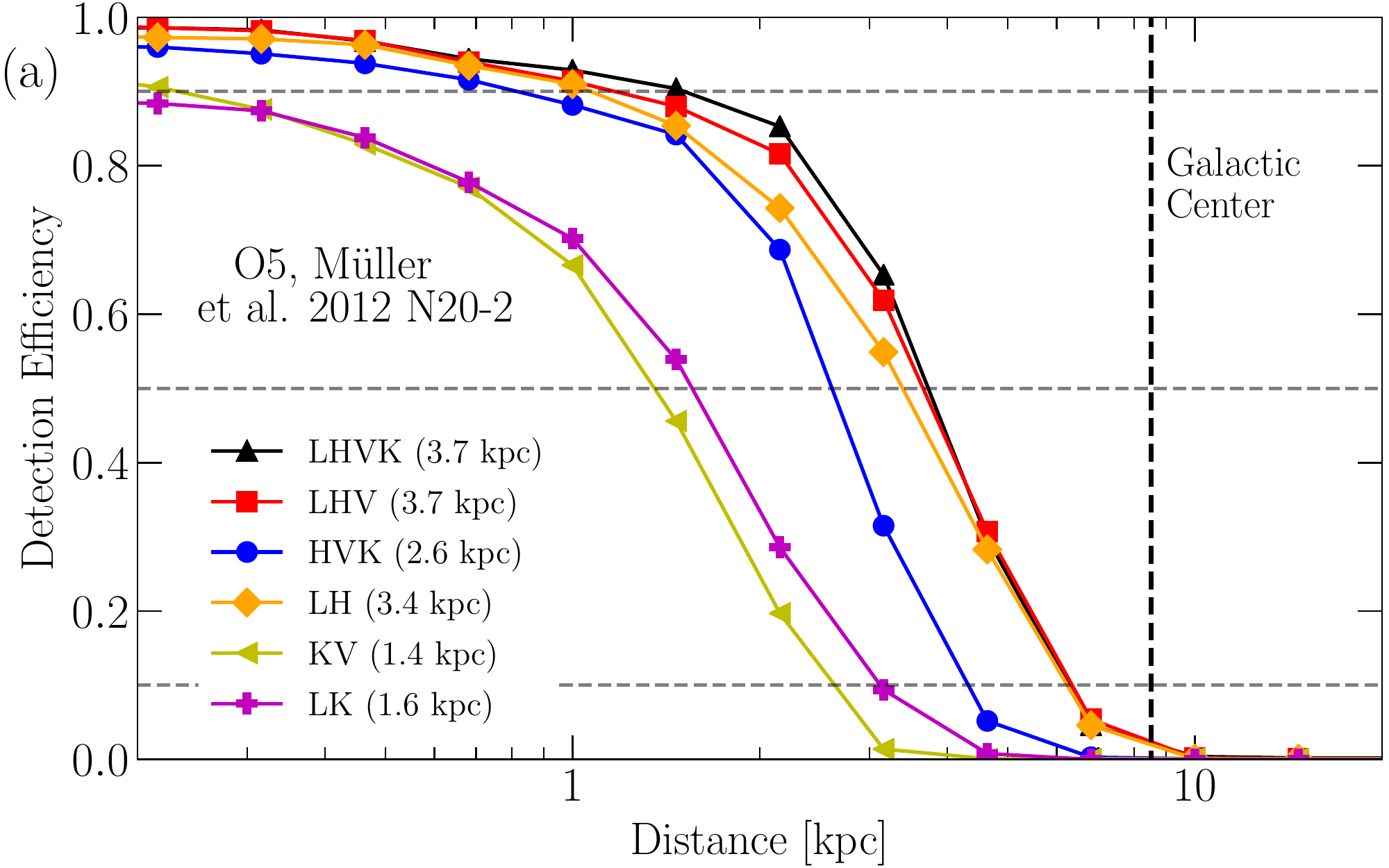}
 \end{minipage}%
 \begin{minipage}[c][][t]{0.495\textwidth}
    \vspace*{\fill}
    \flushright
    \includegraphics[width=0.96\linewidth]{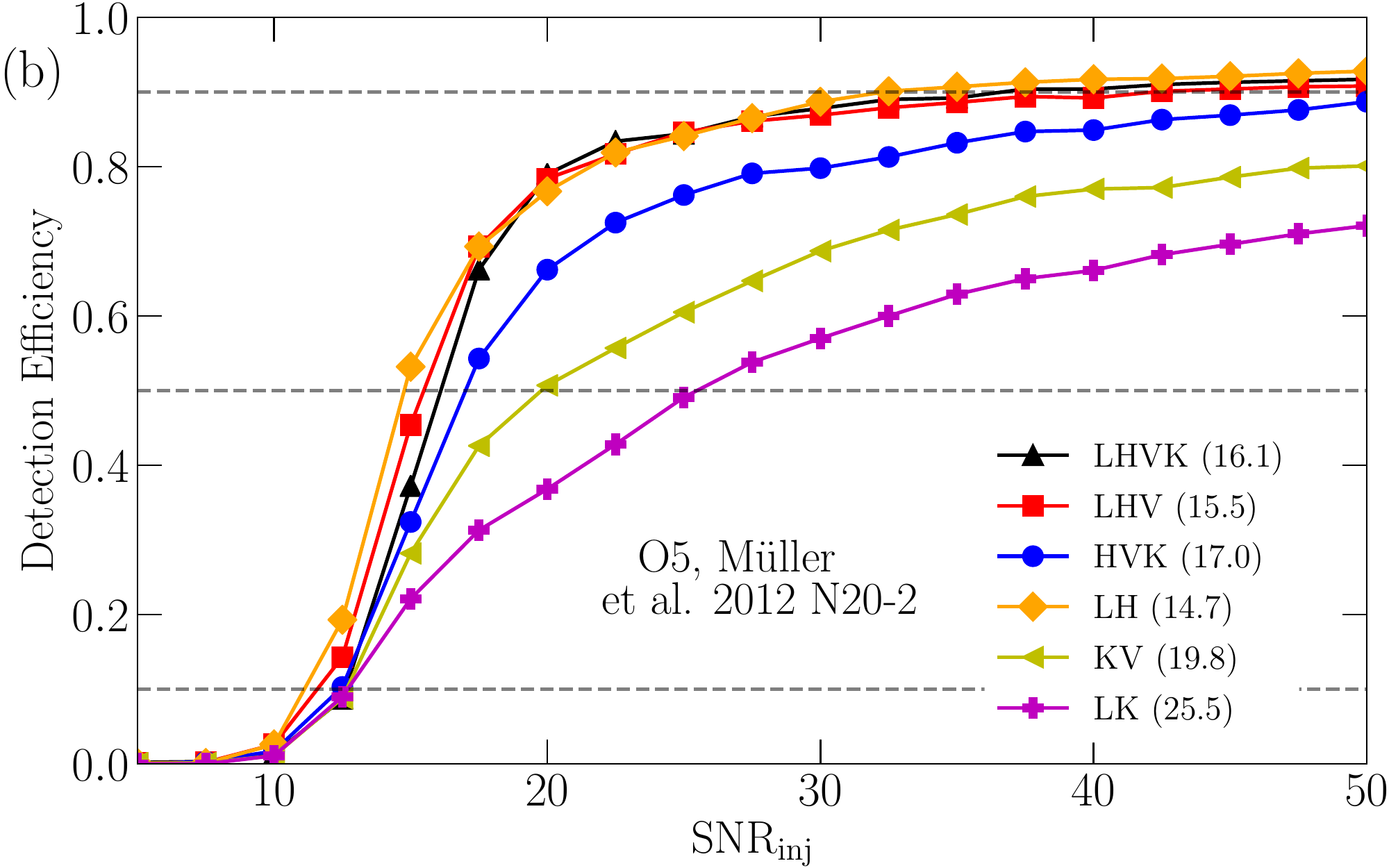}
 \end{minipage}
	\caption{Detection efficiency curves as a function of (a) distance and (b) injected SNR. The numbers in the brackets show detection range and minimum detectable SNR, respectively. The three- and four-detector networks are significantly less sensitive than for two-detector networks.}
	\label{fig:net}
\end{figure*}

\subsection{Inclusion of Virgo and KAGRA}
\label{sec:net}

During O4 and O5 the network of GW detectors will consist of four detectors~\cite{Aasi:2013wya}: L1, H1, V1, and K1. They will have different sensitivities which are depicted in Figure~\ref{fig:hchar}. While adding detectors in the network increases the collected SNR and helps source localization~\cite{Aasi:2013wya}, there are certain challenges, especially when some detectors are less sensitive. It includes optimizing the cWB internal thresholds or designing selection cuts but discussing these challenges is beyond the scope of this study. We simplify our method to provide an approximate estimation of the potential of the four detector network in the context of GW signals from CCSNe. 

For all detector networks, we do not perform background analysis and no selection cuts are applied besides $\rho>6$ (in the previous HL analysis it corresponds roughly to the significance of FAR$<$1/year). For the sensitivity study, the Mul+12 N20-2 waveform is used. Four detectors provide 11 possible network configurations: HLVK, HLV, HVK, LVK, LHK, HL, LV, HV, KV, LK, LK, HK.

Figure~\ref{fig:net} shows the detection efficiency curves as a function of distance and injected SNR for different detector networks. Table~\ref{tab:net} summarizes the detector ranges and minimum detectable SNRs for all detector networks. The networks including H1 and L1 detectors, namely HLVK, HLV LHK, and HL, have comparable detection ranges in corresponding observing runs and the minimum detectable SNR in O4 will be similar to the one in O5. The detection ranges for the other three-detector networks, HVK and LVK will be slightly shorter while for the other two-detector networks, LV, HV, KV, LK, and HK will be two times smaller than for networks including HL. The minimum detectable SNR for the events detected in two-detector networks will be significantly larger.


\begin{table}

\caption{
Detector ranges and minimum detectable SNRs for projected O4 and O5 sensitivities for 11 possible detector configurations. The analysis is performed on the Mul+12 N20-2 waveform.
  }
\begin{tabular*}{\columnwidth} {@{\extracolsep{\textwidth minus \textwidth}} ccccc}
\hline
\hline
\multicolumn{1}{c}{Network}
&\multicolumn{2}{c}{Det. range [kpc]}
&\multicolumn{2}{c}{Min. detect. SNR}
\\
\phantom{0}
& O4 & O5 & O4 & O5 \\
\hline
\hline
HLVK& 2.1 & 3.7 & 15.9 & 16.1 \\
LHV & 1.9 & 3.7 & 15.5 & 15.5 \\
HVK & 1.6 & 2.6 & 16.4 & 17.0 \\
LVK & 1.5 & 2.6 & 16.7 & 17.3 \\
LHK & 2.1 & 3.7 & 14.4 & 14.5 \\
LH  & 1.8 & 3.4 & 14.9 & 14.7 \\
LV  & 0.8 & 1.6 & 27.0 & 25.3 \\
HV  & 0.8 & 1.7 & 25.9 & 24.1 \\
KV  & 0.8 & 1.4 & 19.6 & 19.8 \\
LK  & 1.1 & 1.6 & 20.7 & 25.5 \\
HK  & 1.2 & 1.6 & 19.2 & 25.5 \\
\hline
\hline
\end{tabular*}
\label{tab:net}
\end{table}


\section{Summary}
\label{sec:summary}

Core-collapse supernovae are one of the most spectacular phenomena known in the Universe. CCSN explosions are multi-messenger sources and their emitted GWs are yet to be detected. Although these sources have been modeled for decades, the explosion mechanism and the details of physical processes inside an exploding star are still not fully understood. The detection of these GWs might shed light on rich stochastic dynamics. We analyzed 18 waveform families that represent an extensive set of possible signal morphologies. This wide range of models represents several emission processes, such as prompt-convection, PNS oscillations, SASI/convection, core bounce, and BH formation. The typical GW energy range is from around $10^{-10}\,\mathrm{M}_\odot$ to $10^{-7}\,\mathrm{M}_\odot$ and the peak frequencies range from approximately 100\,Hz to 1\,kHz. 

It is not possible to predict robustly a GW signal emitted by a CCSN, so the search algorithm needs to use weak or minimal assumptions on the signal morphology. Then, using minimal assumptions, we used the coherent WaveBurst algorithm to make predictions on the detectability of the next nearby CCSN event. We predict that in O5, the typical detection range for neutrino-driven explosions will be around 10\,kpc. For models involving rapid rotation of the progenitor stars, the detection range can get up to above 100\,kpc (and possibly more if a strong turbulent GW production continues after the end of the current simulations).

Our analysis of the minimum detectable SNR indicates that the GWs from CCSNe are detectable in the SNR range of roughly 10-25. The shorter waveforms are detectable with smaller SNR, while the longer and broadband signals require larger SNR to be detected. The latter are more challenging to detect, and their reconstructed SNR is usually underestimated.

We quantified the accuracy of the cWB reconstruction using waveform overlap between injected and reconstructed waveforms. The best accuracy is achieved for the short duration signals, like the core bounce. By considering particular GW emission processes, we find that the signals from PNS oscillations require an SNR of 30-40 to become visible in the time-frequency domain. The GW signatures from SASI/convection and prompt-convection can be detectable at SNR values of 15. To capture the full BH formation evolution of 2\,s, in the example considered here, the SNR needs to be around 50.

We analyzed the detectability of GW signals with all possible detector network configurations of LIGO, Virgo, and KAGRA. The distance range and minimum detectable SNR are comparable for four and three detector networks and HL. The detection ranges for two detector networks excluding HL will be around two times smaller.

We conclude that the success of detecting and reconstructing GWs from the next nearby CCSN will depend on several ingredients and in this paper we listed some of the challenges. The algorithms should be prepared before observing the next nearby CCSN and some efforts have already been made. The cWB search may play a significant role in this discovery. One aspect that is not discussed in this paper is the role that the multi-messenger observations could have in increasing the detection confidence and feature reconstruction. We leave this aspect for future publications.


\section{Acknowledgements}

This research has made use of data, software, and/or web tools obtained from the Gravitational Wave Open Science Center, a service of LIGO Laboratory, the LIGO Scientific Collaboration, and the Virgo Collaboration. The work by SK was supported by NSF Grant No. PHY 1806165. MZ was supported by NSF Grant No. PHY-1806885. NS was supported by the Foundation for Polish Science grant TEAM/2016-3/19. NS, DGR, and PS acknowledge support through Polish NCN  grant  no. UMO-2017/26/M/ST9/00978. Mukherjee, Nurbek, and Valdez’s work is supported by NSF PHY-1505861 and NSF PHY-1912630. DGR and PS acknowledge support through the COST Actions CA16104 and  CA16214. DGR  was partially supported by  POMOST/2012-6/11  Program of  Foundation for  Polish  Science co-financed by the  European  Union within the  European  Regional Development  Fund. MC and YZ are partially supported by NSF award PHY-2011334. We gratefully acknowledge the support of LIGO and Virgo for the provision of computational resources. JP is supported by the Australian Research Council (ARC) Discovery Early Career Researcher Award (DECRA) project number DE210101050 and the ARC Centre of Excellence for Gravitational Wave Discovery (OzGrav) project number CE170100004. We thank many authors for accessing their waveforms: Ernazar Abdikamalov, Haakon Andresen, Pablo Cerd\'a-Dur\'an, Harold Dimmelmeier, Takami Kuroda, Anthony Mezzacappa, Viktoriya Morozova, Bernhard M\"uller, Ewald M\"uller, Evan O'Connor, Sean Couch, Christian Ott, David Radice, Sherwood Richers, Simon Scheidegger, Konstantin Yakunin, Matteo Bugli, Martin Obergaulinger, Kuo-Chuan Pan, and others.


\bibliography{library}

\end{document}